\documentclass[a4paper,11pt]{article}
\pdfoutput=1 

\usepackage{jheppub} 
\usepackage{etoolbox}
    \makeatletter
    \patchcmd{\maketitle}{\@fpheader}{}{}{}
    \makeatother                     

\usepackage[T1]{fontenc} 
\usepackage{slashed}

\newcommand{\be}{\begin{equation}}
\newcommand{\ee}{\end{equation}}
\newcommand{\bea}{\begin{eqnarray}}
\newcommand{\eea}{\end{eqnarray}}

\title{\boldmath Leptoquark Option for $B$-meson Anomalies and Leptonic Signatures}

\preprint{}

\author[]{Hyun Min Lee}

\affiliation[]{Department of Physics, Chung-Ang University, Seoul 06974, Korea}

\emailAdd{hminlee@cau.ac.kr }

\abstract{We entertain the option of scalar leptoquarks to explain the anomalies in the semi-leptonic decays of $B$-mesons and the discrepancies in the lepton $(g-2)_l$'s including the recent results at Fermilab E989.   The $R_{K^{(*)}}$ and $R_{D^{(*)}}$ anomalies can be accommodated by the specific couplings for triplet and singlet leptoquarks, respectively, subject to the bounds from $B\to K\nu{\bar\nu}$. We discuss the correlation between the leptonic signatures from leptoquarks such as $\mu\to e\gamma$ and the electric dipole moment of electron and show that desirable neutrino masses can be generated dominantly by top-quark loops in the extension of the model with several doublet leptoquarks. 
}

\begin{document} 
\maketitle
\flushbottom

\section{Introduction}
\label{sec:intro}

The universality of fundamental interactions of leptons, apart from the Higgs boson interactions, the so called Lepton Flavor Universality (LFU), has been well tested within the Standard Model (SM), so any deviation from LFU would hint at new physics beyond the SM. Recent anomalies from the semi-leptonic decays of $B$-mesons, the so called $R_{K^{(*)}}$ \cite{RK,RK-new,RK-update,RK-belle-new,RKs,RKs-new,P5} and $R_{D^{(*)}}$ anomalies \cite{lhcb,babar,belle,belle-new}, favor the modified interactions of muon and tau leptons beyond the SM, respectively, although we still need to understand the hadronic uncertainties in related B-meson decays \cite{qcd}.

The measurements of the anomalous magnetic moments $(g-2)_l$ and/or electric dipole moments  $d_l$ for leptons are also important tests of the SM at one-loop level, given that the SM predictions and the measured values for them are in good agreement up to high precision. However, the measurement of the magnetic dipole moment of muon at Brookhaven E821 has shown a $3.7\sigma$ deviation from the SM  \cite{amu-exp,amu} and it has been confirmed at $3.3\sigma$ by Fermilab $g-2$ collaboration \cite{fermilab}, resulting in the combined significance at $4.2\sigma$. There is a similar deviation observed in the case of the electron counterpart although at a smaller significance of $2.4\sigma$ \cite{ae-exp,ae}.

In this article, we consider the extension of the SM with singlet and triplet scalar leptoquarks to explain both $R_{K^{(*)}}$ and $R_{D^{(*)}}$ anomalies simultaneously, being consistent with other bounds from $B$-meson observables and flavor constraints.  We update the constraints on the leptoquark couplings in view of the latest results from LHCb and Belle experiments. 

We present the flavor structure for leptoquark couplings \cite{LQs,leptoquarks,LQ-hmlee,LQ-g2,LQ-fit} violating the LFU for the updated results for the $B$-meson decays \cite{newfit,newfit2} and accommodating the discrepancies in the magnetic moments of muon and electron \cite{LQ-hmlee,LQ-g2,charmquark}, while satisfying the bounds from the flavor violating decays of leptons. We consider two benchmark models for leptoquark couplings: one is the minimal scenario to explain the $B$-meson anomalies and account for the $(g-2)_\mu$ anomaly with top-quark loops at the same time, and another is to explain the $B$-meson anomalies as well as the $(g-2)_\mu$ and $(g-2)_e$ anomalies via top-quark and charm-quark loops, respectively.  In either case, we discuss the correlation between the leptoquark couplings required for the $B$-meson anomalies and the leptonic signatures including the electric dipole moment of electron and show how neutrino masses are generated at loops in the model.

The paper is organized as follows. 
We begin with a review on the $B$-meson anomalies and identify the parameter space for explaining them with singlet and triplet scalar leptoquarks. We also present a general discussion on the bounds from $B\to K{\nu} {\bar\nu}$ and the magnetic and electric dipole moments of leptons and the flavor-violating decays of leptons in this model. We give the details on the benchmark models for the leptoquark couplings and show the correlations between the $B$-meson anomalies and $(g-2)_l$ anomalies and other leptonic signatures. Finally, conclusions are drawn.

\section{Leptoquarks for $B$-meson anomalies}

We first give an overview on the experimental status of the measurements on rare B-meson decays, $R_{K^{(*)}}$ and  $R_{D^{(*)}}$ and discuss models with $SU(2)_L$ singlet and triplet scalar leptoquarks to explain $R_{D^{(*)}}$ and $R_{K^{(*)}}$ anomalies, respectively \cite{leptoquarks,LQ-hmlee,LQ-g2,LQ-fit}. 
Then, we update the phenomenological conditions for the $B$-meson anomalies.

\subsection{Overview on the status of B-meson anomalies}

In this section, we give a brief overview on the status of the $B$-meson anomalies and the interpretations in terms of the effective Hamiltonians in the SM will be made in the next subsections.

The reported values of $R_K={\cal B}(B\rightarrow K\mu^+\mu^-)/{\cal
  B}(B\rightarrow Ke^+e^-)$ from LHCb 2011-2012 \cite{RK} and 2015-2016 data  \cite{RK-new} were deviated from the SM prediction by $2.6\sigma$ and $2.5\sigma$, respectively.
Recently, the result has been updated further with LHCb 2017-2018 data \cite{RK-update}, so the combined value for $1.1\,{\rm GeV}^2<q^2<6.0\,{\rm GeV}^2$ now shows a $3.1\sigma$ deviation from the SM, as follows,
\bea
R_K=0.846^{+0.042}_{-0.039}({\rm stat})^{+0.013}_{-0.012}({\rm syst}).
\eea
The analysis of the full Belle data sample led to new results on $R_K$ in various bin energies \cite{RK-belle-new}, showing the consistency with  the LHCb result  in the bin of interest, $1\,{\rm GeV}^2<q^2 <6\,{\rm GeV}^2$.

On the other hand for vector $B$-mesons, $R_{K^*}={\cal
  B}(B\rightarrow K^*\mu^+\mu^-)/{\cal B}(B\rightarrow
K^*e^+e^-)$ from LHCb~\cite{RKs} is
\bea
  R_{K^*}= \left\{ \begin{array}{cc} 0.66^{+0.11}_{-0.07}({\rm stat})\pm 0.03({\rm syst}), \quad 0.045\,{\rm GeV}^2<q^2 <1.1\,{\rm GeV}^2, \vspace{0.2cm} \\  0.69^{+0.11}_{-0.07}({\rm stat})\pm 0.05({\rm syst}), \quad 1.1\,{\rm GeV}^2<q^2 <6.0\,{\rm GeV}^2, \end{array}\right.
\eea
which again differs from the SM prediction by 2.1--$2.3\sigma$ and
2.4--$2.5\sigma$, depending on the energy bins. The deviation in $R_{K^*}$ is supported by the reduction in the angular distribution of $B\rightarrow K^*\mu^+\mu^-$, the so called $P'_5$ variable~\cite{P5}. There is also a recent update on $R_{K^*}$ from Belle data \cite{RKs-new}, also showing a similar deviation in particular in low energy bins ($0.045\,{\rm GeV}<q^2<1.1\,{\rm GeV}$), although the errors are still large.

Taking the results of BaBar \cite{babar}, Belle \cite{belle} and LHCb \cite{lhcb} for $R_D={\cal B}(B\rightarrow D\tau\nu)/{\cal  B}(B\rightarrow Dl\nu)$ and $R_{D^*}={\cal B}(B\rightarrow D^*\tau\nu)/{\cal  B}(B\rightarrow D^*l\nu)$ with $l=e,\mu$ for BaBar and Belle and $l=\mu$ for LHCb, the Heavy Flavor Averaging Group \cite{hflav} reported the experimental world averages as follows,
\bea
R^{\rm exp}_D &=& 0.403\pm 0.040\pm 0.024, \\
R^{\rm exp}_{D^*}&=& 0.310\pm 0.015\pm 0.008.
\eea
On the other hand, taking into account the lattice calculation of $R_D$, which is $R_D=0.299\pm 0.011$ \cite{RD-lattice}, and the uncertainties in $R_{D^*}$ in various groups \cite{RD-SM,RD-SMetc}, we take the SM predictions for these ratios as follows,
\bea
R^{\rm SM}_D &=& 0.299\pm 0.011, \\
R^{\rm SM}_{D^*}&=& 0.260\pm 0.010.
\eea
Then, the combined derivation between the measurements and the SM predictions for $R_D$ and $R_{D^*}$ is about $4.1\sigma$. 
We quote the best fit values for $R_D$ and $R_{D^*}$ including the new physics contributions \cite{RD-bestfit},
\bea
\frac{R_D}{R^{\rm SM}_D}=\frac{R_{D^*}}{R^{\rm SM}_{D^*}}=1.21\pm 0.06. \label{RDratio}
\eea 
There has been a recent update on $R_{D^{*}}$ from Belle \cite{belle-new}, which leads to the new global average values, 
$R^{\rm exp}_D = 0.337\pm 0.030$ and $R^{\rm exp}_{D^*}= 0.299\pm 0.013$.

The upcoming LHCb with Run 3 and HL-LHC data will measure the rates of semi-leptonic B-meson decays with better precision and  the online Belle II experiment can eventually test the lepton flavor universality in $B$-meson decays up to a few percent level  at least with data of $5\,{\rm ab}^{-1}$ \cite{belle2}.

\subsection{Effective interactions from scalar leptoquarks}

We consider the Lagrangian for an $SU(2)_L$ singlet scalar leptoquark $S_1$ with $Y=+\frac{1}{3}$, and an $SU(2)_L$ triplet scalar leptoquark, $S_3\equiv \Phi_{ab}$ with $Y=+\frac{1}{3}$, as follows,
\bea
{\cal L}_{LQ}= {\cal L}_{S_1}+{\cal L}_{S_3} \label{leptolag}
\eea
with
\bea
{\cal L}_{S_1} =-\lambda_{ij} \overline{(Q^C)^a_{Ri}}\, (i\sigma^2)_{ab} \,S_1 \, L^b_{Lj} - \lambda'_{ij} \overline{(u^C)_{Li}}S_1 e_{jR} +{\rm h.c.} \label{sLQ}
\eea
where $a,b$ are $SU(2)_L$ indices, $\sigma^2$ is the second Pauli matrix and $\psi^C=C{\bar\psi}^T$ is the charge conjugate with $C=i\gamma^0\gamma^2$,
and
\bea
{\cal L}_{S_3} =-\kappa_{ij} \overline{(Q^C)^a_{Ri}}\, \Phi_{ab}\, L^b_{Lj} +{\rm h.c.} \label{tLQ}
\eea
with
\be
\Phi_{ab}=\left( \begin{array}{cc} \sqrt{2}\phi_3 & -\phi_2 \\  -\phi_2  & -\sqrt{2} \phi_1 \end{array} \right)
\ee
where $(\phi_1,\phi_2,\phi_3)$ forms an isospin triplet with $T_3=+1,0,-1$ and $Q=+\frac{4}{3}, +\frac{1}{3}, -\frac{2}{3}$.  We note that our conventions can be compared to those in the literature from $\Phi=(i\sigma^2) ({\vec \sigma}\cdot {\vec S})$ where $\vec\sigma$ are Pauli matrices and $\vec S$ are complex scalar fields. 

It is worthwhile to mention that the flavor mixing in the SM should be taken into account for identifying the necessary leptoquark couplings for $B$-meson decays. First, in the basis where the mass matrices for down-type quarks and charged leptons are diagonal, from eq.~(\ref{leptolag}), we can rewrite the general leptoquark couplings for mass eigenstates (denoted by primed fields), as follows,
 \bea
 {\cal L}_{LQ} &=&-\lambda^{(u)}_{ij}  \overline{(u^{\prime C})_{Ri}} \,S_1 \, e'_{Lj} +\lambda^{(d)}_{ij}  \overline{(d^{\prime C})_{Ri}} \,S_1 \, \nu'_{Lj} - y'_{ij} \overline{(u^{\prime C})_{Li}}S_1 e'_{jR}  \nonumber \\
 &&+\sqrt{2} {\tilde \kappa}_{1,ij}  \overline{(d^{\prime C})_{Ri}}\,\phi_1\, e'_{jL} +{\tilde \kappa}^{(u)}_{2,ij}  \overline{(u^{\prime C})_{Ri}}\,\phi_2\, e'_{jL}+{\tilde \kappa}^{(d)}_{2,ij}  \overline{(d^{\prime C})_{Ri}}\,\phi_2 \,\nu'_{jL}  \nonumber \\
 &&-\sqrt{2}{\tilde \kappa}_{3,ij}  \overline{(u^{\prime C})_{Ri}}\,\phi_3\, \nu'_{jL}  \label{masseigen}
 \eea
 where
 \bea
 \lambda^{(u)}&=& (V_{\rm CKM})^* \cdot \lambda, \,\,\,\, \lambda^{(d)}= \lambda \cdot U^\dagger_{\rm PMNS},  \, \,\,\, y'=V^*_{u_R} \cdot \lambda', \label{dbasis1}\\ 
{\tilde\kappa}_1&=&\kappa, \,\,\,\,  {\tilde \kappa}^{(u)}_{2} = (V_{\rm CKM})^*\cdot \kappa,\,\,\,\, {\tilde \kappa}^{(d)}_{2} = \kappa\cdot U^\dagger_{PMNS}, \,\, \,\,{\tilde \kappa}_{3}=(V_{\rm CKM})^*\cdot\kappa\cdot U^\dagger_{\rm PMNS}.\label{dbasis2} 
 \eea
 Here, $V_{\rm CKM}, U_{\rm PMNS}, V_{u_R}$ are the CKM matrix for quarks, the PMNS matrix for leptons, and the rotation matrix for right-handed up-type quarks, respectively. Then, at least, for $\lambda^{(u)}$ and ${\tilde\kappa}_1$ couplings, which are relevant for $R_{D^{(*)}}$ and $R_{K^{(*)}}$ anomalies, respectively, we can infer the original flavor structure for leptoquarks.  For instance, if $\lambda_{33}\neq 0$ and the other components of $\lambda_{ij}$ are zero, we can get the hierarchical leptoquark couplings, $\lambda^{(u)}_{33}=V_{tb}\lambda_{33}\simeq  \lambda_{33}$, $\lambda^{(u)}_{23}=V_{cb}\lambda_{33}\simeq 0.041 \lambda_{33}$, and $\lambda^{(u)}_{13}=V_{ub} \lambda_{33}\simeq 0.0036\lambda_{33}$.

Instead, taking the basis where the mass matrices for up-type quarks and charged leptons are diagonal, from eq.~(\ref{leptolag}), we can also rewrite the general leptoquark couplings for mass eigenstates in eq.~(\ref{masseigen}) with
\bea
 \lambda^{(u)}&=& \lambda, \,\,\,\, \lambda^{(d)}=(V_{\rm CKM})^T\cdot \lambda \cdot U^\dagger_{\rm PMNS},  \, \,\,\, y'=\lambda',  \label{ubasis1} \\ 
{\tilde\kappa}_1&=&(V_{\rm CKM})^T\cdot\kappa, \,\,\,\,  {\tilde \kappa}^{(u)}_{2} = \kappa,\,\,\,\, {\tilde \kappa}^{(d)}_{2} = (V_{\rm CKM})^T\cdot\kappa\cdot U^\dagger_{PMNS}, \,\, \,\,{\tilde \kappa}_{3}=\kappa\cdot U^\dagger_{\rm PMNS}.  \label{ubasis2}
\eea
In this case, similarly to the previous case with the diagonal mass matrix for down-type quarks, the flavor structure for leptoquarks are inherited in $\lambda^{(u)}$ and ${\tilde\kappa}_1$ couplings up to the CKM mixing, which are relevant for $R_{D^{(*)}}$ and $R_{K^{(*)}}$ anomalies, respectively. More importantly, there is no CKM mixing appearing in  $\lambda^{(u)}$ and $y'=\lambda'$, which are relevant for explaining $(g-2)_l$ anomalies, as will be discussed in the later section. For convenience, we work in the mass eigenstates for the leptoquark couplings in the following discussion.

We also remark that other Yukawa couplings for the singlet leptoquark are possible, such as $\lambda^{\prime\prime}_{ij}{\bar d}_{Ri} S_1 (u^c)_{Lj}$, but the simultaneous presence of $\lambda_{ij}\neq 0$ and $\lambda^{\prime\prime}_{ij}\neq 0$ would be dangerous for proton decay, as in $R$-parity violating supersymmetry. So, we set $\lambda^{\prime\prime}_{ij}=0$ in favor of explaining the $(g-2)_\mu$ anomaly.  But, if we keep $\lambda^{\prime\prime}_{ij}\neq 0$ instead, it would also lead to an interesting phenomenology for leptoquark searches with dijets. In $R$-parity violating supersymmetry, the Yukawa couplings to the singlet quarks and leptons, $\lambda'_{ij}$, are absent, but the non-holomorphic terms, $\frac{1}{M^2_*}\int d^2\theta d^2{\bar\theta} \, c_{ij} X (U^c)^\dagger_i S_1 (E^c)^\dagger_j +{\rm h.c.}$, with $M_*$ being the cutoff scale and $c_{ij}$ being dimensionless parameters, could generate $\lambda'_{ij}= c_{ij}\frac{F_X}{M^2_*}$ for $X=F_X\theta^2$. In this case,  down-type squarks corresponding to $S_1$ are responsible for both $R_{D^{(*)}}$ and $(g-2)_\mu$ anomalies \cite{LQ-hmlee}.

After integrating out the leptoquark scalars in the Lagrangian (\ref{leptolag}), we obtain the effective Lagrangian for the SM fermions in the following,
\bea
{\cal L}_{\rm eff} &=& \Bigg(\frac{1}{4m^2_{S_1}}\, \lambda_{ij}\lambda^*_{kl}+\frac{3}{4m^2_{S_3}}\, \kappa_{ij} \kappa^*_{kl} \Bigg) \Big({\bar Q}_{Lk}\gamma^\mu Q_{Li}\Big)\Big({\bar L}_{Ll}\gamma_\mu L_{Lj}\Big) \nonumber \\
&&+ \Bigg(-\frac{1}{4m^2_{S_1}}\, \lambda_{ij}\lambda^*_{kl}+\frac{1}{4m^2_{S_3}}\, \kappa_{ij} \kappa^*_{kl} \Bigg) \Big({\bar Q}_{Lk}\gamma^\mu\sigma^I Q_{Li}\Big)\Big({\bar L}_{Ll}\gamma_\mu \sigma^I L_{Lj}\Big) \nonumber \\
&&-\frac{1}{2m^2_{S_1}}\,\lambda_{ij} \lambda^{\prime *}_{kl}\left(\Big({\bar u}_{Rk}Q_{Lj}\Big)\Big({\bar e}_{Rl}L_{Lj}\Big)-\frac{1}{4}\Big({\bar u}_{Rk}\sigma^{\mu\nu}Q_{Li}\Big)\Big({\bar e}_{Rl}\sigma_{\mu\nu}L_{Lj}\Big) \right)+{\rm h.c.} \nonumber \\
&&+\frac{1}{2m^2_{S_1}}\, \lambda^\prime_{ij} \lambda^{\prime *}_{kl}\Big({\bar u}_{Rk}\gamma^\mu u_{Ri}\Big)\Big({\bar e}_{Rl}\gamma_\mu e_{Rj}\Big) \label{LQeff}
\eea
where  $\sigma^I(I=1,2,3)$ are the Pauli matrices.
There, we find that there are both $SU(2)_L$ singlet and triplet $V-A$ operators from the leptoquark couplings to left-handed fermions in the SM.
Moreover, for $\lambda^\prime_{ij}\neq 0$, which is crucial to explain the muon $(g-2)_\mu$, the decay mode, $B\rightarrow D\tau \nu_\tau$, can be affected by additional scalar and tensor operators on top of the effective operators obtained in Ref.~\cite{LQ-hmlee}, as shown in eq.~(\ref{LQeff}). A simultaneous solution to $R_{K}$ and $R_{D^{(*)}}$ anomalies was previously discussed along the line of the effective Lagrangian approach \cite{Bdecayeff}, and it was also pointed out that leptoquark couplings are important in relation to $b\to c\tau {\bar \nu}$ decays in the literature \cite{LQ-RD}.

\subsection{$R_{D^{(*)}}$ anomalies and leptoquarks}

The effective Hamiltonian for $b\rightarrow c  \tau \nu$ in the SM is given by
\bea
{\cal H}_{\rm eff}&=&\frac{4G_F}{\sqrt{2}} V_{cb} \bigg[ (1+C_{VL})\, ({\bar c}\gamma^\mu P_L b) ({\bar\tau}\gamma_\mu P_L \nu_\tau)+C_{VR}\, ({\bar c}\gamma^\mu P_R b) ({\bar\tau}\gamma_\mu P_L \nu_\tau)+C_{SL}\, ({\bar c} P_L b) ({\bar\tau} P_L \nu_\tau) \nonumber \\
&&+C_{SR}\, ({\bar c} P_R b) ({\bar\tau} P_L \nu_\tau)+C_T ({\bar c}\,\sigma^{\mu\nu} P_L b) ({\bar\tau}\sigma_{\mu\nu} P_L \nu_\tau)\bigg] +{\rm h.c.} \label{RDeff}
\eea
where  $C_{VL}=0$ and $C_{VR}=C_{SL}=C_{SR}=C_T=0$ in the SM and $V_{cb}\approx 0.04$.  The new physics contribution may also contain the other dimension-6 four-fermion vector operators that are not present in the SM. In order to explain the $R_{D^{(*)}}$ anomalies in eq.~(\ref{RDratio}), however, the Wilson coefficients for the new physics contribution are favored to be  $C_{VL}=0.1$ and $C_{VR}=C_{SL}=C_{SR}=C_T=0$  in eq.~(\ref{RDeff}), while taking $[0.072,0.127]$ and $[0.044,0.153]$ for $C_V$ within $1\sigma$ and $2\sigma$ errors.

From the part of the effective Lagrangian obtained in eq.~(\ref{LQeff}) after the leptoquark $S_1$ is integrated, we take the effective Hamiltonian relevant for $b\rightarrow c\tau{\bar\nu}_\tau$ as
\bea
{\cal H}^{S_1}_{b\rightarrow c\tau{\bar\nu}_\tau}
&=&-\frac{\lambda^*_{33}\lambda_{23}}{2 m^2_{S_1}}\, ({\bar b}_L \gamma^\mu c_L) ({\bar\nu}_{\tau L}\gamma_\mu \tau_L) +{\rm h.c.} \nonumber \\
&&-\frac{\lambda^*_{23}\lambda'_{33}}{2 m^2_{S_1}} \bigg[({\bar c} P_L b) ({\bar\tau} P_L \nu_\tau)-\frac{1}{4} ({\bar c}\,\sigma^{\mu\nu} P_L b) ({\bar\tau}\sigma_{\mu\nu} P_L \nu_\tau) \bigg]+{\rm h.c.} \nonumber \\
&\equiv&  \frac{1}{\Lambda^2_D} \, ({\bar b}_L \gamma^\mu c_L) ({\bar\nu}_{\tau L}\gamma_\mu \tau_L) +{\rm h.c.} \nonumber \\
&&+\frac{1}{\Lambda^{\prime 2}_D} \, \bigg[({\bar c} P_L b) ({\bar\tau} P_L \nu_\tau)-\frac{1}{4} ({\bar c}\,\sigma^{\mu\nu} P_L b) ({\bar\tau}\sigma_{\mu\nu} P_L \nu_\tau) \bigg]+{\rm h.c.}.  \label{Deff}
\eea
As a consequence, the singlet leptoquark gives rise to the effective operator for explaining the $R_{D^{(*)}}$ anomalies and the effective cutoff scales are to be $\Lambda_D\sim 3.5\,{\rm TeV}\ll \Lambda'_D$. Thus, for $m_{S_1}\gtrsim 1\,{\rm TeV}$, we need $\sqrt{\lambda^*_{33}\lambda_{23}}\gtrsim 0.4$.

\begin{figure}
  \begin{center}
    \includegraphics[height=0.45\textwidth]{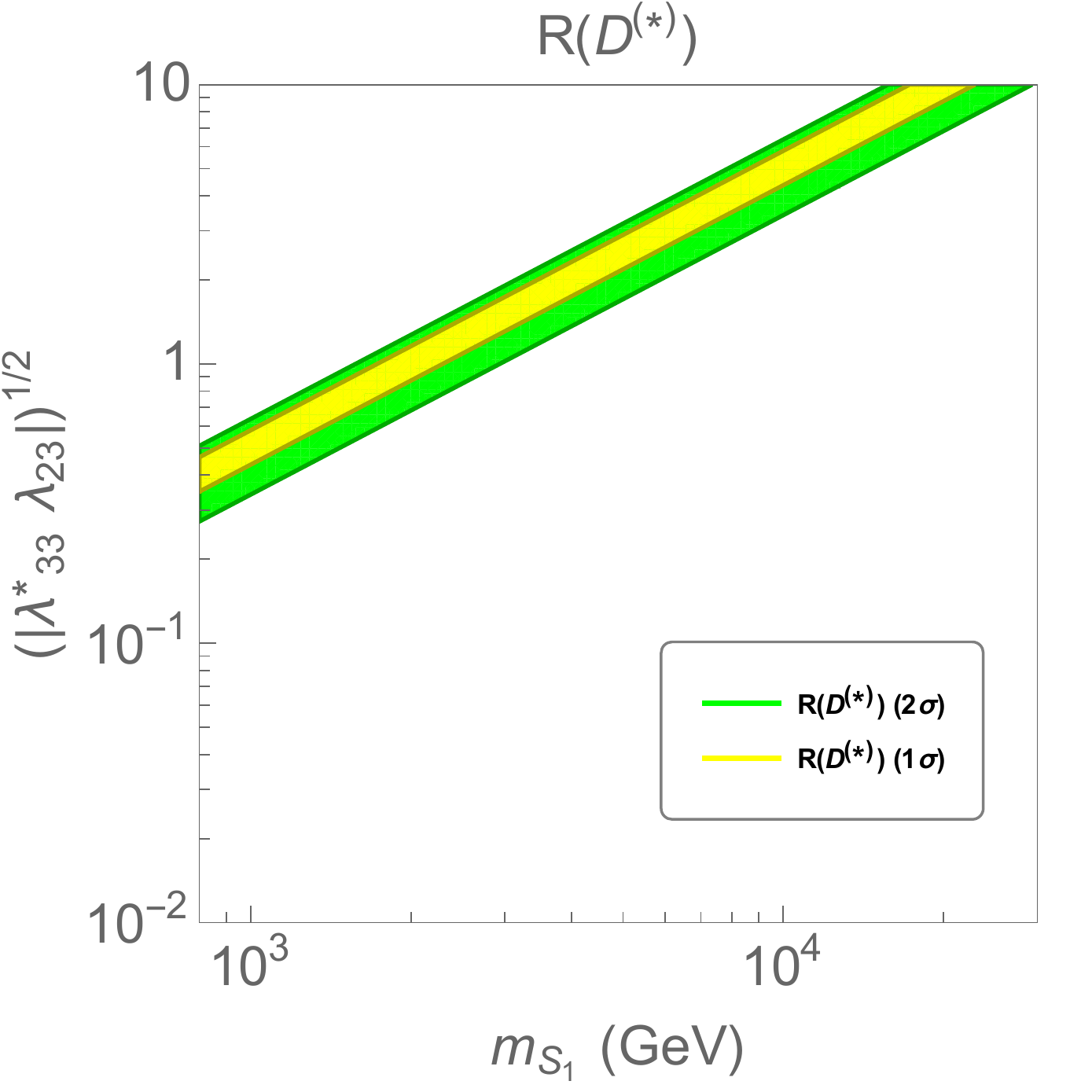}\,\,
     \includegraphics[height=0.45\textwidth]{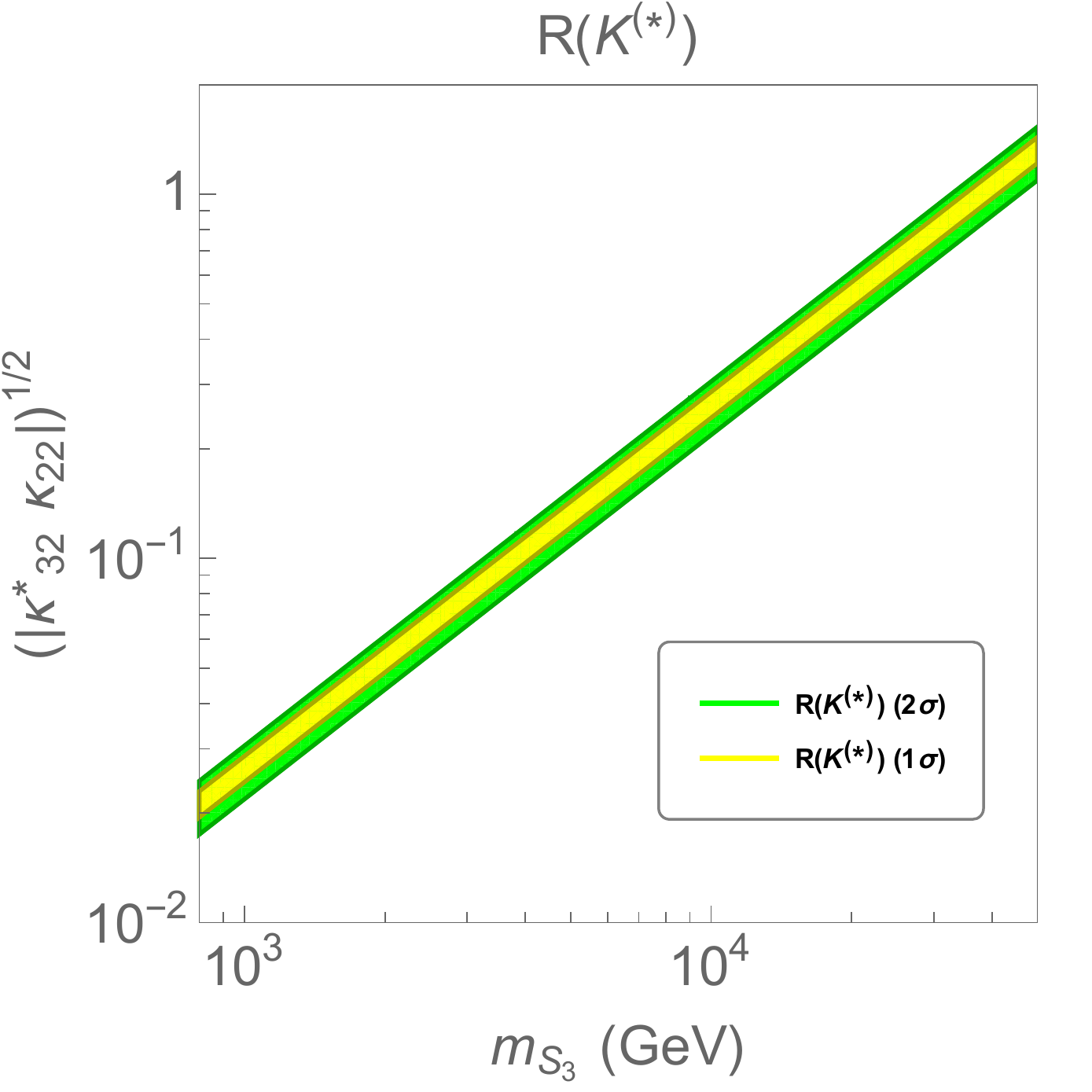}
      \end{center}
  \caption{Parameter space  for the leptoquark mass $m_{LQ}$ and the effective coupling $\lambda_{\rm eff}$, explaining the B-meson anomalies, in green(yellow) region at $2\sigma(1\sigma)$ level. We have taken $m_{LQ}=m_{S_1}$ and $\lambda_{\rm eff}=\sqrt{|\lambda^*_{33}\lambda_{23}|}$ for $R_{D^{(*)}}$ on left plot, and $m_{LQ}=m_{S_3}$ and $\lambda_{\rm eff}=\sqrt{|\kappa^*_{32}\kappa_{22}|}$ for $R_{K^{(*)}}$ on right plot.  }
  \label{B-anomalies}
\end{figure}

In the left plot of Fig.~\ref{B-anomalies}, we depict the parameter space for $m_{S_1}$ and the effective leptoquark coupling, $\lambda_{\rm eff}=\sqrt{|\lambda^*_{33}\lambda_{23}|}$, in which the $R_{D^{(*)}}$ anomalies can be explained within $2\sigma(1\sigma)$ errors  in green(yellow) region from the conditions below eq.~(\ref{RDeff}).

\subsection{$R_{K^{(*)}}$ anomalies and leptoquarks}

The effective Hamiltonian for $b\rightarrow s  \mu^+ \mu^-$ is given by
\bea
\Delta {\cal H}_{{\rm eff},{\bar b}\rightarrow {\bar s}\mu^+ \mu^-} = -\frac{4G_F}{\sqrt{2}}  \,V^*_{ts} V_{tb}\,\frac{\alpha_{em}}{4\pi}\, (C^{\mu}_9 {\cal O}^\mu_9+C^{\mu}_{10} {\cal O}^\mu_{10}+C^{\prime\mu}_9{\cal O}^{\prime\mu}_9+C^{\prime\mu}_{10}{\cal O}^{\prime\mu}_{10} )+{\rm h.c.} \label{RKeff}
\eea
where $ {\cal O}^\mu_9 \equiv ({\bar s}\gamma^\mu P_L b) ({\bar \mu}\gamma_\mu \mu)$, $ {\cal O}^\mu_{10} \equiv ({\bar s}\gamma^\mu P_L b) ({\bar \mu}\gamma_\mu\gamma^5 \mu)$, $ {\cal O}^{\prime\mu}_9 \equiv ({\bar s}\gamma^\mu P_R b) ({\bar \mu}\gamma_\mu \mu)$ and $ {\cal O}^{\prime\mu}_{10} \equiv ({\bar s}\gamma^\mu P_R b) ({\bar \mu}\gamma_\mu\gamma^5 \mu)$, and $\alpha_{\rm em}$ is the electromagnetic coupling. In the SM, the Wilson coefficients are given by $C^{\mu,\rm SM}_9(m_b)=-C^{\mu,\rm SM}_{10}(m_b)=4.27$ and $C^{\prime\mu, {\rm SM}}_9(m_b)\approx-C^{\prime\mu,{\rm SM}}_{10}(m_b)\approx 0$. 

For $C^{\mu,\rm NP}_{10}=C^{\prime\mu, {\rm NP}}_9=C^{\prime\mu,{\rm NP}}_{10}=0$, the best-fit value for new physics contribution is given by $C^{\mu, {\rm NP}}_9=-0.82\pm 0.14$ \cite{newfit} to explain all the rare $B$ decays including the $R_{K^{(*)}}$ anomalies.
On the other hand, for $C^{\mu, {\rm NP}}_9=-C^{\mu,\rm NP}_{10}$ and others being zero, the best-fit value for new physics contribution is given by $C^{\mu, {\rm NP}}_9=-0.43\pm 0.07$ \cite{newfit}.

From the part of the effective Lagrangian obtained in eq.~(\ref{LQeff}) after the component of the triplet leptoquark, $\phi_1$ with $Q=+\frac{4}{3}$, is integrated out,   we also take the effective Hamiltonian relevant for $b\rightarrow s\mu^+\mu^-$ as
\bea
{\cal H}^{S_3}_{b\rightarrow s\mu^+\mu^-}
=-\frac{\kappa^*_{32}\kappa_{22}}{ m^2_{\phi_1}}\, ({\bar b}_L \gamma^\mu s_L) ({\bar\mu}_L\gamma_\mu \mu_L) +{\rm h.c.}\equiv \frac{1}{\Lambda^2_K}\, ({\bar b}_L \gamma^\mu s_L) ({\bar\mu}_L\gamma_\mu \mu_L) +{\rm h.c.}.
\eea
As a consequence, the triplet leptoquark gives rise to the effective operator of the $(V-A)$ form for the quark current, that is, $C^{\mu,\rm NP}_9=-C^{\mu,\rm NP}_{10}\neq 0$, as favored by the $R_{K^{(*)}}$ anomalies, and the effective cutoff scale is to be $\Lambda_K\sim 30\,{\rm TeV}$. The result is in contrast to the case for $Z'$ models with family-dependent charges such as $Q'=x(B_3-L_3)+y(L_\mu-L_\tau)$ with $x, y$ being arbitrary parameters  where $C^{\mu,\rm NP}_9\neq 0$ and $C^{\mu,\rm NP}_{10}=0$ \cite{Zprime}. Then, for $m_{\phi_1}\gtrsim 1\,{\rm TeV}$, we need $\sqrt{\kappa^*_{32}\kappa_{22}}\gtrsim 0.03$.  
Therefore, we can combine scalar leptoquarks, $S_1$ and $S_3$, to explain $R_{D^{(*)}}$ and $R_{K^{(*)}}$ anomalies, respectively. 

In the right plot of Fig.~\ref{B-anomalies}, we depict the parameter space for $m_{S_3}$ and the effective leptoquark coupling, $\kappa_{\rm eff}=\sqrt{|\kappa^*_{32}\kappa_{22}|}$, in which the $R_{K^{(*)}}$ anomalies can be explained within $2\sigma(1\sigma)$ errors in green(yellow) region from the conditions below eq.~(\ref{RKeff}).
There are direct limits on the masses and couplings of scalar leptoquarks from the LHC, corresponding to $m_{\rm LQ}\gtrsim 1\,{\rm TeV}$ from most of leptoquark decay channels \cite{LQ-hmlee}.

\section{Other bounds on leptoquark couplings}

In this section, we consider the bounds from other $B$-meson decays such as $B\rightarrow K^{(*)}\nu {\bar \nu}$ and present the general results for the magnetic and electric dipole moments of leptons due to scalar leptoquarks.

\subsection{Rare meson decays and mixing}

In leptoquark models, there is no $B-{\bar B}$ mixing at tree level. The new contribution of leptoquarks to the $B_s-{\bar B}_s$ mixing appears sizable at one-loop level in the case of the singlet leptoquark \cite{leptoquarks}, but it is less constrained due to large errors in the SM prediction. 
However, both singlet and triplet leptoquarks contribute to $B\rightarrow K^{(*)}\nu {\bar \nu}$ at tree level, which are important constraints for leptoquark models. 

The effective Hamiltonian relevant for ${\bar b}\rightarrow {\bar s}\nu {\bar \nu}$ \cite{BKnunu} is given by
\bea
{\cal H}_{{\bar b}\rightarrow {\bar s}\nu {\bar \nu}}= -\frac{\sqrt{2}\alpha_{\rm em} G_F}{\pi}\, V_{tb} V^*_{ts} \sum_l C^l_L  ({\bar b}\gamma^\mu P_L s)({\bar \nu}_{l}\gamma_\mu P_L\nu_{l}) 
\eea
where $C^l_L=C^{\rm SM}_L+C^{l,{\rm NP}}_\nu$. Here, the SM contribution $C^{\rm SM}_L$ is given by $C^{\rm SM}_L=-X_t/s^2_W$
with $s_W\equiv \sin\theta_W$ and $X_t=1.469\pm 0.017$.
From the result in eq.~(\ref{LQeff}), the new contributions from scalar leptoquarks to the effective Hamiltonian for $B\rightarrow K\nu {\bar \nu}$ are
\bea
C^{l,{\rm NP}}_\nu= -\left(  \frac{\lambda^*_{3i}\lambda_{2j}}{2m^2_{S_1}}+\frac{\kappa^*_{3i} \kappa_{2j}}{2m^2_{\phi_2}}\right) \, \frac{\pi}{ \sqrt{2}\alpha_{\rm em} G_FV_{tb} V^*_{ts}}. 
\eea
Then, the ratio of the branching ratios in our model is given by
\bea
R_{K^{(*)}\nu}&\equiv& \frac{B(B\rightarrow K^{(*)}\nu{\bar\nu})}{B(B\rightarrow K^{(*)}\nu{\bar\nu})\Big|_{\rm SM}} \nonumber \\
&=& \frac{2}{3}+\frac{1}{3} \frac{|C^{\rm SM}_L+C^{l,{\rm NP}}_\nu|^2}{|C^{\rm SM}_L|^2}.
\eea
The experimental bounds on $B(B\rightarrow K^{(*)}\nu{\bar\nu})$ \cite{BKnunu-exp} are given by
\bea
B(B\rightarrow K\nu{\bar\nu})<1.6\times 10^{-5}, \quad\quad B(B\rightarrow K^{*}\nu{\bar\nu})< 2.7\times 10^{-5},
\eea
and  the SM predictions \cite{BKnunu-SM} are
\bea
B(B\rightarrow K\nu{\bar\nu})\Big|_{\rm SM}&=&(3.98\pm 0.43\pm 0.19)\times 10^{-6}, \nonumber   \\
\quad  B(B\rightarrow K^{*}\nu{\bar\nu})\Big|_{\rm SM}&=&(9.19\pm 0.86\pm 0.50)\times 10^{-6}.
\eea
As a result, the $R_{K^*\nu}$ bound on the new physics contribution is
\bea
-10.1<{\rm Re}(C^{l,{\rm NP}}_\nu)<22.8.
\eea

Taking into account $\kappa_{32}$ and $\kappa_{22}$, which are necessary for $B\rightarrow K^{(*)}\mu^+\mu^-$, the triplet scalar leptoquark contributes only to $B\rightarrow K^{(*)}\nu_\mu{\bar\nu}_\mu$.
In this case, as the triplet leptoquark contribution to $C^{\mu,{\rm NP}}_\nu$ is about the same as $C^{\mu,{\rm NP}}_9=-0.61$,  it  satisfies the $R_{K^*\nu}$ bound on its own easily. However, it is necessary to introduce $\lambda_{33}$ and $\lambda_{23}$ for the singlet leptoquark for explaining the anpmalies in $B\rightarrow D^{(*)}\tau{\bar\nu}_\tau$, resulting in the significant contribution to $B\rightarrow K^{(*)}\nu_\tau {\bar \nu}_\tau$.

Ignoring the mass splitting within the triplet scalar leptoquark, we get $m_{\phi_1}=m_{\phi_2}=m_{\phi_3}\equiv m_{S_3}$. Then, we can cancel the contributions to  $B\rightarrow K^{(*)}\nu_\tau{\bar\nu}_\tau$ or $B\rightarrow K^{(*)}\nu_{\mu,\tau}{\bar\nu}_{\tau,\mu}$, with the additional couplings, $\kappa_{23}$ and $\kappa_{33}$,  for the triplet leptoquark, as follows,
\bea
\frac{|\kappa^*_{33} \kappa_{23}|}{|\lambda^*_{33} \lambda_{23}|} \approx  \frac{|\kappa^*_{32}\kappa_{23}|}{|\lambda^*_{32}\lambda_{23}| } \approx \frac{m^2_{S_3}}{m^2_{S_1}}. \label{lam32}
\eea
Then, for $m_{S_3}\sim m_{S_1}$, we need $\sqrt{|\kappa^*_{33}\kappa_{23}|}\approx\sqrt{ |\lambda^*_{33} \lambda_{23}|}\gtrsim 0.4$, in order to explain the $R_{D^{(*)}}$ anomalies.
On the other hand,  for $m_{S_3}\sim m_{S_1}$, the additional coupling for the singlet leptoquark, $\lambda_{32}$, must satisfy $\sqrt{|\lambda^*_{32} \lambda_{23}|}\approx\sqrt{|\kappa^*_{32}\kappa_{23}|}$, subject to the conditions, $\sqrt{\lambda^*_{33}\lambda_{23}}\gtrsim 0.4$ and $\sqrt{|\kappa^*_{32}\kappa_{22}|}\gtrsim 0.03$, for explaining $R_{D^{(*)}}$ and  $R_{K^{(*)}}$ anomalies, respectively. In this case,  it is possible to get a sizable $\lambda_{32}$ coupling in order to explain the deviation in $(g-2)_\mu$  with top-quark loops as will be discussed in the next subsection.

We also remark that the singlet leptoquark loops contribute to $(g-2)_e$ via charm-quark loops for $\lambda_{21}\neq 0$ and $\lambda'_{21}\neq 0$, instead of top-quark loops. In this case, we also need to consider the constraints from $R_{K^*\nu}$, which again sets the similar relations between the leptoquark couplings as in eq.~(\ref{lam32}), as follows,
\bea
\frac{|\kappa^*_{33} \kappa_{21}|}{|\lambda^*_{33} \lambda_{21}|} \approx  \frac{|\kappa^*_{32}\kappa_{21}|}{|\lambda^*_{32}\lambda_{21}| } \approx \frac{m^2_{S_3}}{m^2_{S_1}}  \label{lam21}
\eea
Therefore, we also need to take into account a nonzero $\kappa_{21}$ for a sizable $\lambda_{21}$ coupling.

Finally, we note that the leptoquark couplings required for $R_{D^{(*)}}$ anomalies also lead to a nonzero extra contribution to $B_c\rightarrow \tau {\bar\nu}_\tau$, so the corresponding decay branching ratio \cite{Btaunu1} becomes
\bea
{\rm Br}(B_c\rightarrow \tau {\bar\nu}_\tau) =0.02 \bigg(\frac{f_{B_c}}{0.43\,{\rm GeV}} \bigg)^2 \Big| 1+C_{VL}-4.3 C_{SL} \Big|^2 
\eea
where the Wilson coefficients are given in eq.~(\ref{RDeff}) with eq.~(\ref{Deff}).
The best limit for $B_c\rightarrow \tau {\bar\nu}_\tau$ \cite{Btaunu2,Btaunu1} is given by
\bea
{\rm Br}(B_c\rightarrow \tau {\bar\nu}_\tau) \leq 0.1.
\eea
Therefore, choosing $C_{SL}\ll  C_{VL}\simeq 0.1$ to explain the $R_{D^{(*)}}$ anomalies in our model as discussed in Section 2.3, we find that the bound from $B_c\rightarrow \tau {\bar\nu}_\tau$ does not constrain the parameter space strongly in our model.

\subsection{Magnetic and electric dipole moments of leptons}

The deviation of the anomalous magnetic moment of muon between experiment and SM
values is given \cite{amu-exp,amu} by
\be
\Delta a_\mu = a^{\rm exp}_\mu -a^{\rm SM}_\mu = 279(76)\times 10^{-11}, \label{amu-dev}
\ee
which is a $3.7\sigma$ discrepancy from the SM. Recently, the muon $g-2$ experiment E989 at Fermilab has confirmed it by a $3.3\sigma$ deviation from the SM \cite{fermilab}.
As a result, from the combined average with Brookhaven E821, the difference from the SM value  becomes
\bea
\Delta a_\mu = a^{\rm exp}_\mu -a^{\rm SM}_\mu =251(59)\times 10^{-11}, \label{amu-recent}
\eea
which is now a $4.2\sigma$  discrepancy from the SM \cite{fermilab}.

Furthermore, there is a $2.4\sigma$ discrepancy reported between the SM prediction for the anomalous magnetic moment of electron and the experimental measurements \cite{ae-exp,ae}, as follows,
\bea
\Delta a_e = a^{\rm exp}_e -a^{\rm SM}_e =-88(36)\times 10^{-14}. \label{ae-dev}
\eea
There have been some attempts to explain both anomalies beyond the SM \cite{aemodels}.

From the Yukawa couplings for the singlet scalar leptoquark, the chirality-enhanced effect from the top quark contributes most to the $(g-2)_l$ of leptons \cite{leptoquarks}, as follows,
\bea
a^{S_1}_l= \frac{m_l }{4\pi^2}\,{\rm Re}[C^{ll}_R] \label{g2}
\eea
with
\bea
C^{ij}_R\equiv -\frac{N_c}{12m_{S_1}^2} \,\sum_k m_{u_k} \lambda_{ki} \lambda^{\prime *}_{kj} \Big(7+4\log \Big(\frac{m^2_{u_k}}{m^2_{S_1}} \Big) \Big).
\eea

On the other hand, a new source of CP violation in the singlet leptoquark couplings could lead to a new contribution to the electric dipole moment (EDM) of lepton. 
\bea
d_l=  \frac{1 }{8\pi^2}\,{\rm Im}[C^{ll}_R]. \label{edm}
\eea
The current best limit on the electron EDM comes from ACMEII  \cite{acme}  as follows,
 \bea
 d_e<1.1\times 10^{-29}\,{\rm e\,cm}. 
 \eea
Moreover, we have a much weaker bound on the muon EDM \cite{muedm} by
\bea
d_\mu<1.5\times 10^{-19}\,{\rm e\,cm}. 
\eea

\subsection{Flavor violating decays of leptons}

The leptoquark couplings for muon or electron anomalous magnetic moments also contribute to the branching ratio of the flavor violating decays of tau lepton, $\tau\rightarrow \mu\gamma$ and $\tau\rightarrow e\gamma$, as follows,
\bea
{\rm BR}(\tau\rightarrow \mu\gamma)&=& \frac{\alpha m^3_\tau}{256\pi^4}\, \tau_\tau \Big(|C^{23}_R|^2+|C^{23}_L|^2 \Big), \\ \label{taumu}
{\rm BR}(\tau\rightarrow e\gamma)&=& \frac{\alpha m^3_\tau}{256\pi^4}\, \tau_\tau \Big(|C^{13}_R|^2+|C^{13}_L|^2 \Big) \label{taue}
\eea
where $C^{ij}_L=C^{ij}_R( \lambda_{3i}\rightarrow \lambda^{\prime }_{3i}, \lambda'_{3j}\rightarrow \lambda_{3j} )$ and the lifetime of tau is given by $\tau_\tau=(290.3\pm 0.5)\times 10^{-15}\, {\rm s}$ \cite{pdg}. 
The current experimental bounds are given \cite{taumu} by 
\bea
{\rm BR}(\tau\rightarrow\mu\gamma)&<&4.4\times 10^{-8}, \\
{\rm BR}(\tau\rightarrow e\gamma)&<&3.3\times 10^{-8}.
 \eea
 
 Similarly, the simultaneous presence of contributions to both muon and electron anomalous magnetic moments generically leads to the branching ratio of $\mu\rightarrow e \gamma$,
 \bea
{\rm BR}(\mu\rightarrow e\gamma)= \frac{\alpha m^3_\mu}{256\pi^4}\, \tau_\mu \Big(|C^{12}_R|^2+|C^{12}_L|^2 \Big) \label{mue}
\eea
where the lifetime of muon is given by $\tau_\mu=2.197\times 10^{-6}\, {\rm s}$ \cite{pdg}. 
The current experimental bound is given \cite{meg} by 
\be
{\rm BR}(\mu\rightarrow e \gamma)<4.2\times 10^{-13}.
 \ee

\section{Leptoquark flavors and leptonic signatures}

We discuss two benchmark models for leptoquarks, Model I and Model II, in light of the $B$-meson anomalies, and show the correlations with the leptonic signatures from the leptoquark couplings, such as the magnetic and electric dipole moments of leptons, flavor violating decays of leptons and neutrino masses.

In this section, we focus on the case where the singlet leptoquark gives rise to dominant contributions to $(g-2)_l$ anomalies at one-loop level, although it is also possible to use the triplet leptoquark couplings in the presence of a mass mixing between singlet and triplet leptoquarks \cite{charmquark}.

\subsection{Model I: Top-quark loops}

We consider the minimal flavor structure for leptoquark couplings to accommodate both $R_{K^{*}}$  and  $R_{D^{*}}$ anomalies, as follows \cite{LQ-hmlee},
\bea
\lambda=\left(\begin{array}{ccc}  0  & 0 & 0 \\   0 &  0 & \lambda_{23} \\ \lambda_{31} & \lambda_{32} & \lambda_{33}  \end{array} \right), \quad \kappa=\left(\begin{array}{ccc}  0  & 0 & 0 \\   0 &  \kappa_{22} & \kappa_{23} \\ 0 & \kappa_{32} & \kappa_{33}  \end{array} \right), \quad \lambda=\left(\begin{array}{ccc}  0  & 0 & 0 \\   0 &  0 & 0 \\  \lambda'_{31} & \lambda'_{32} & 0  \end{array} \right). \label{flavor1}
\eea
We discuss the reason for the above flavor structure in detail. 
First, we need $\lambda_{33}\lambda_{23}\neq 0$ for $R_{K^{*}}$ anomalies and $\kappa_{32}\kappa_{22}\neq 0$ for  $R_{D^{*}}$ anomalies.
As discussed for eq.~(\ref{lam32}), we also introduced extra couplings for the triplet leptoquark, $\kappa_{23}$ and $\kappa_{33}$, to get a sizable $\lambda_{32}$  for generating a large deviation in $(g-2)_\mu$ without a conflict to the bound from $B\rightarrow K^{(*)}\nu{\bar \nu}$.  We also took $ \lambda'_{33}=0$ to suppress the leptoquark contributions to scalar and tensor operators for $R_{D^{*}}$ in eq.~(\ref{Deff}). 
 We note that $\lambda_{31}, \lambda'_{31}$ are unnecessary for $B$-meson anomalies, but we introduced them unlike in Ref.~\cite{LQ-hmlee} to see new contributions to $(g-2)_e$ and $d_e$ from top-quak loops.

 \begin{figure}
  \begin{center}
    \includegraphics[height=0.45\textwidth]{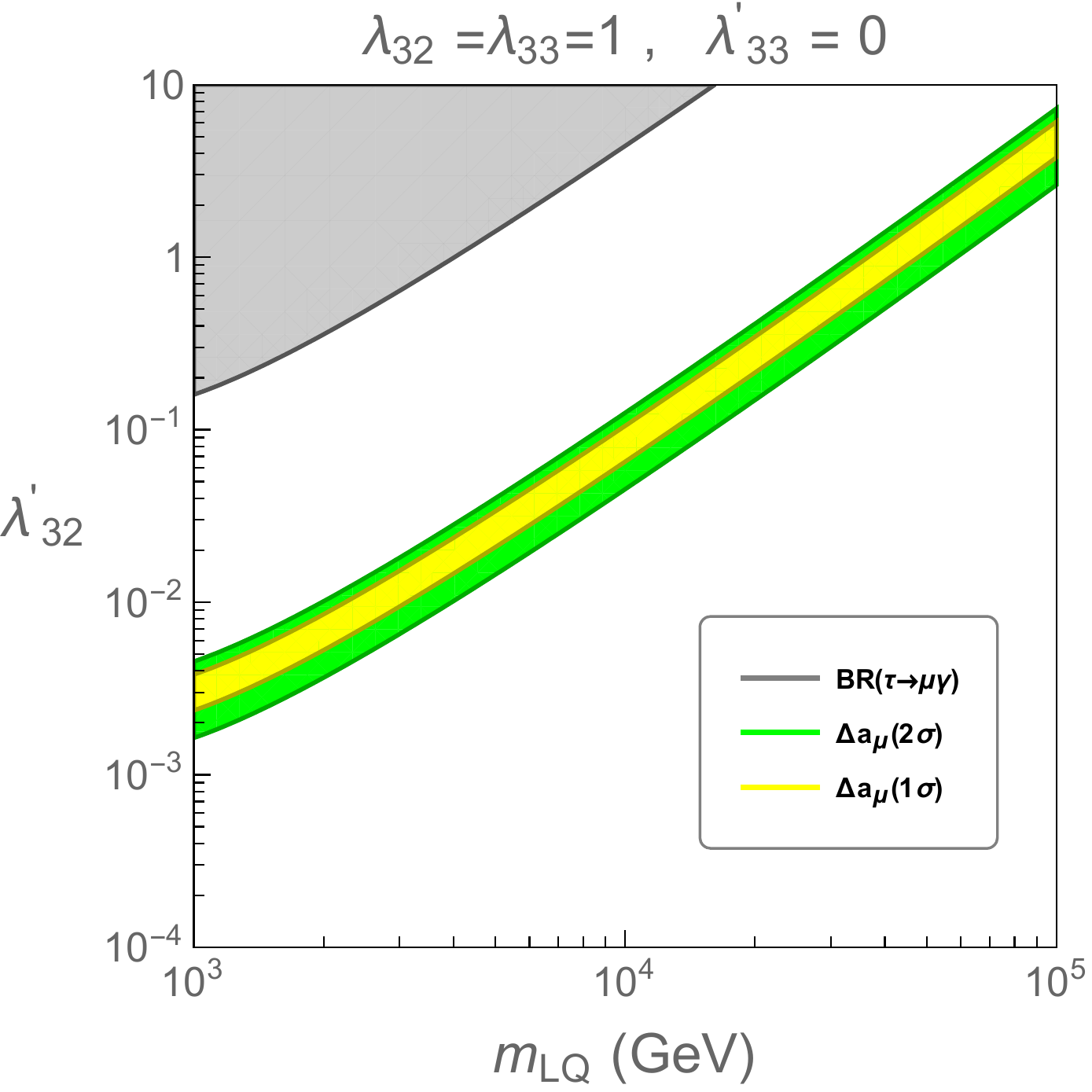}\,\,
     \includegraphics[height=0.45\textwidth]{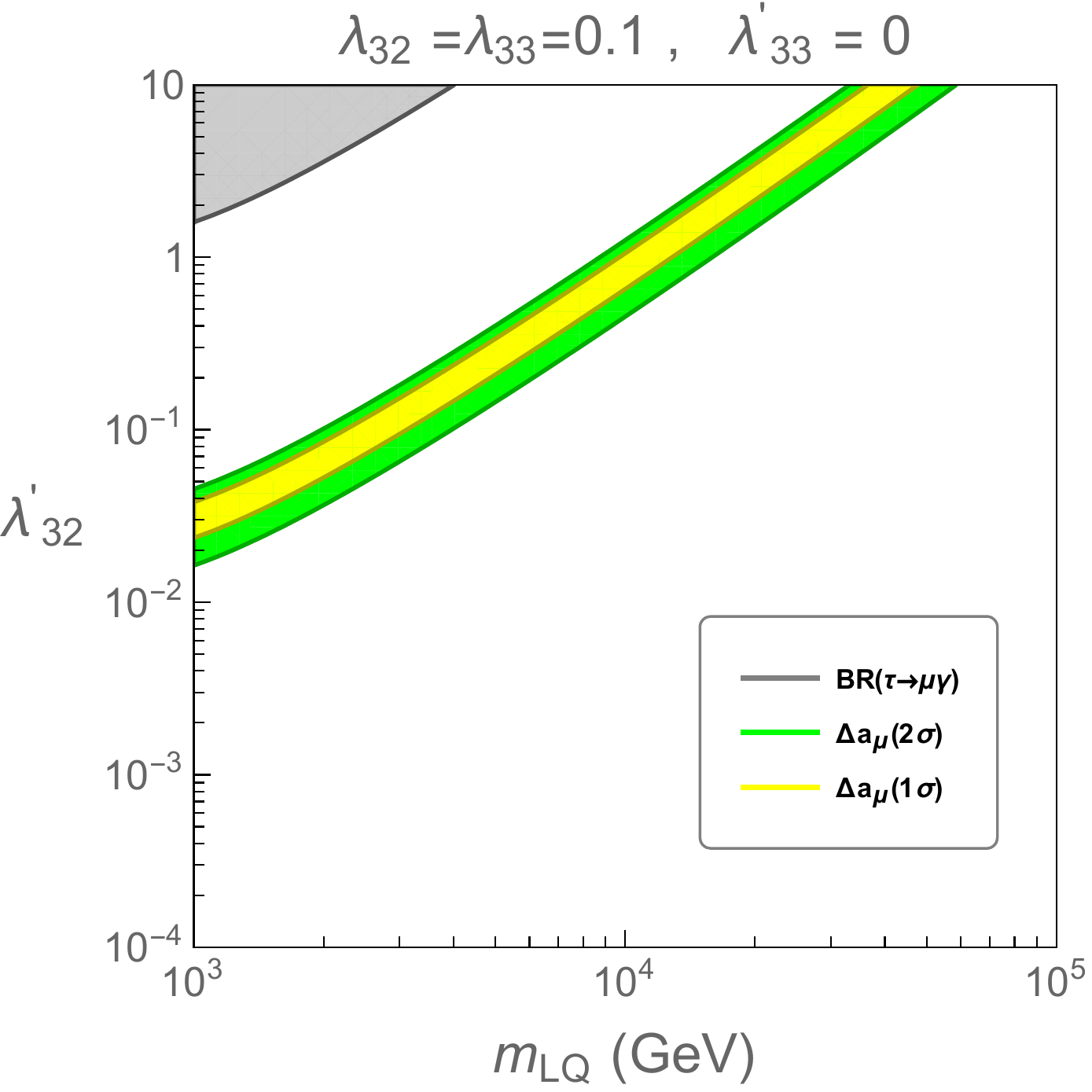}
      \end{center}
  \caption{Parameter space  for $m_{LQ}=m_{S_1}$ and $\lambda'_{32}$ allowed by $(g-2)_\mu$ in Model I. Green(yellow) regions show $2\sigma(1\sigma)$ bands for  $(g-2)_\mu$. The gray region is excluded by the bound on ${\rm BR}(\tau\rightarrow\mu\gamma)$. We have fixed $\lambda_{32}=\lambda_{33}=1(0.1)$ on left(right) plot and $\lambda'_{33}=0$ in both plots.}
  \label{mg2}
\end{figure}

For the flavor structure in eq.~(\ref{flavor1}), from eq.~(\ref{g2}), the one-loop contributions with top quark and singlet leptoquark to the $(g-2)_l$'s for muon and electron, are, respectively, 
\bea
a^{S_1}_\mu &=& \frac{C_t m_\mu m_t}{16\pi^2 m^2_{S_1}} \,{\rm Re}(\lambda_{32}\lambda^{\prime *}_{32}),  \label{amu} \\
a^{S_1}_e &=& \frac{C_t m_e m_t}{16\pi^2 m^2_{S_1}} \,{\rm Re}(\lambda_{31}\lambda^{\prime *}_{31}), \label{ae}
\eea 
with $C_t=4\log \Big(\frac{m^2_{S_1}} {m^2_{t}}\Big)-7 $. Then, in this case, there is a top mass enhancement \cite{LQ-hmlee} for both $(g-2)_\mu$ and $(g-2)_e$, and the ratio of $(g-2)_l$'s depend on the leptoquark couplings by
\bea
\frac{a^{S_1}_e}{a^{S_1}_\mu }= \frac{m_e}{m_\mu}\,\cdot \frac{{\rm Re}(\lambda_{31}\lambda^{\prime *}_{31}) }{{\rm Re}(\lambda_{32}\lambda^{\prime *}_{32})}.
\eea
Then, for $|a^{S_1}_e/a^{S_1}_\mu|\simeq 3\times 10^{-4} $ at the central values in the deviations in eqs.~(\ref{amu-dev}) and (\ref{ae-dev}), we would need $|{\rm Re}(\lambda_{31}\lambda^{\prime *}_{31})/{\rm Re}(\lambda_{32}\lambda^{\prime *}_{32})|\simeq 0.06$. But, there is a  strong bound from $\mu\to e\gamma$, which excludes the possibility to explain the $(g-2)_e$ in this case.

 \begin{figure}
  \begin{center}
    \includegraphics[height=0.45\textwidth]{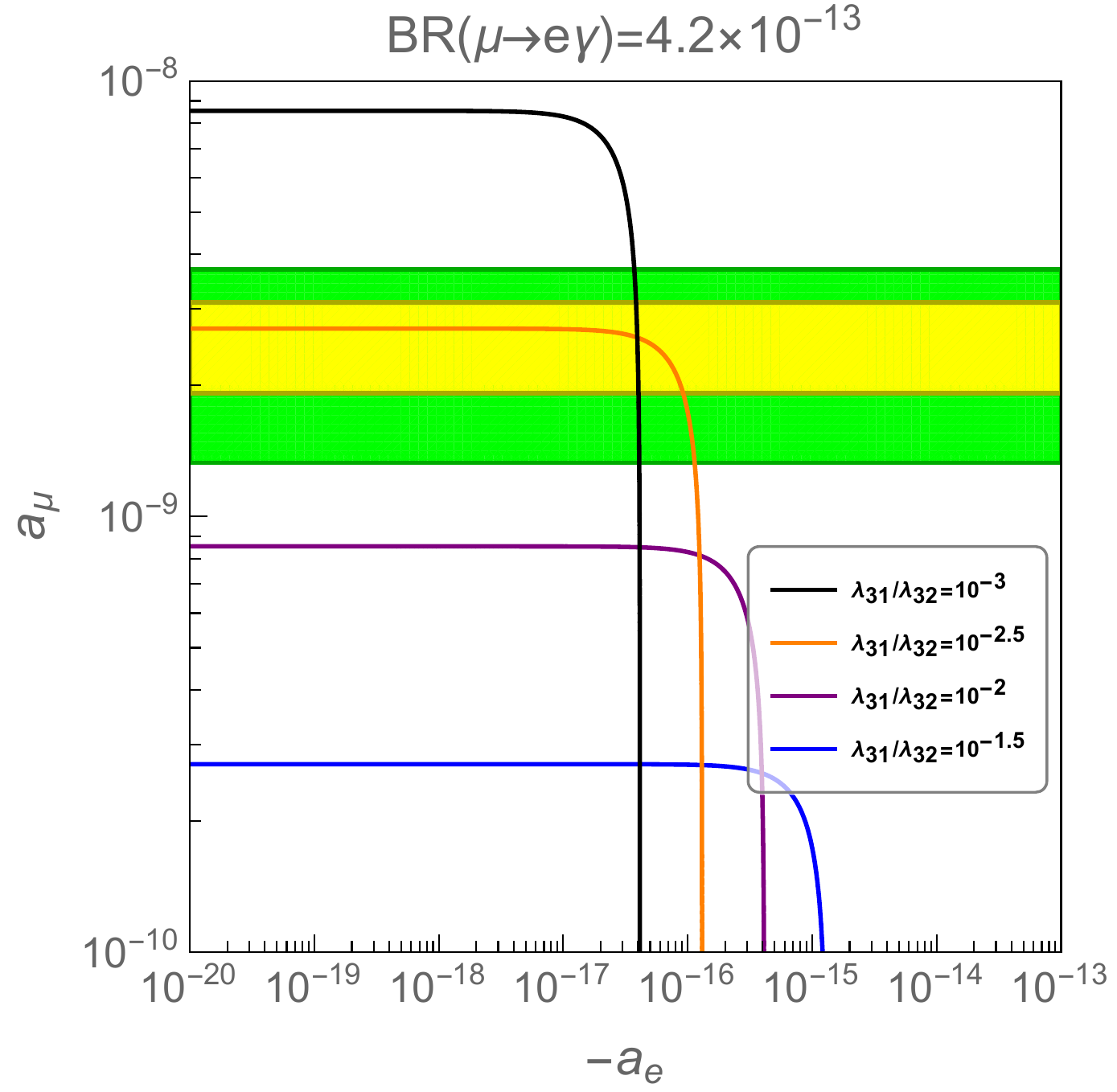}\,\,
     \includegraphics[height=0.45\textwidth]{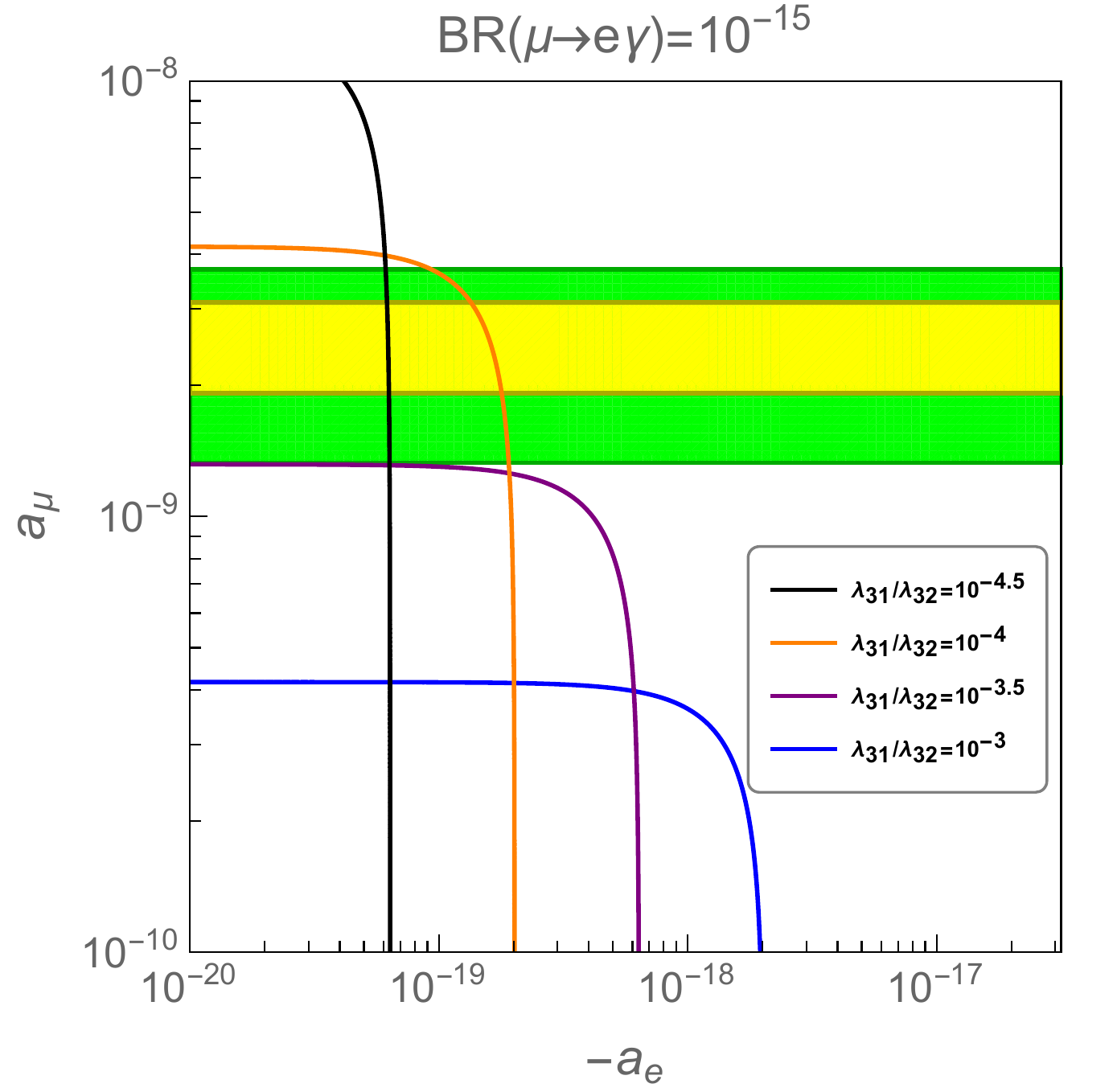}
      \end{center}
  \caption{Correlation between the leptoquark contributions to the muon and electron $g-2$ in Model I.  Green(yellow) regions show $2\sigma(1\sigma)$ bands for  $(g-2)_\mu$. We fixed  ${\rm BR}(\mu\to e\gamma)=4.2\times 10^{-13}, 10^{-15}$ on left and right plots, respectively. Various lines are shown for $\lambda_{31}/\lambda_{32}=10^{-3}, 10^{-2.5}, 10^{-2}, 10^{-1.5}$ from top to bottom (black, orange, purple, blue lines) on left plot and $\lambda_{31}/\lambda_{32}=10^{-4.5}, 10^{-4}, 10^{-3.5}, 10^{-3}$ from top to bottom  (black, orange, purple, blue lines) on right plot. }
  \label{corr1}
\end{figure}

Similarly, from eq.~(\ref{g2}), the lepton EDMs are given by
\bea
d_\mu &=&  \frac{C_t  m_t}{32\pi^2 m^2_{S_1}} \,{\rm Im}(\lambda_{32}\lambda^{\prime *}_{32}), \\
d_e &=& \frac{C_t m_t}{32\pi^2 m^2_{S_1}} \,{\rm Im}(\lambda_{31}\lambda^{\prime *}_{31}).
\eea

On the other hand, from eqs.~(\ref{taumu}) and (\ref{mue}), we obtain the branching ratios for flavor violating decays of leptons,
\bea
{\rm BR}(\tau\rightarrow \mu\gamma)&=& \frac{\alpha m^3_\tau}{256\pi^4}\, \tau_\tau \bigg(\frac{ C_t m_t}{4 m^2_{S_1}} \bigg)^2\,\Big[|\lambda_{32}\lambda^{\prime *}_{33}|^2+|\lambda_{33}\lambda^{\prime *}_{32}|^2 \Big],  \label{BRtaumu} \\
{\rm BR}(\tau\rightarrow e\gamma)&=& \frac{\alpha m^3_\tau}{256\pi^4}\, \tau_\tau \bigg(\frac{ C_t m_t}{4 m^2_{S_1}} \bigg)^2\,\Big[|\lambda_{31}\lambda^{\prime *}_{33}|^2+|\lambda_{33}\lambda^{\prime *}_{31}|^2 \Big],  \label{BRtaue} \\
{\rm BR}(\mu\rightarrow e \gamma)&=& \frac{\alpha m^3_\mu}{256\pi^4}\, \tau_\mu \bigg(\frac{ C_t m_t}{4 m^2_{S_1}} \bigg)^2\,\Big[|\lambda_{31}\lambda^{\prime *}_{32}|^2+|\lambda_{32}\lambda^{\prime *}_{31}|^2 \Big].
\eea
Assuming that the leptoquark couplings are real,  we can eliminate $\lambda_{32}$ and $\lambda_{31}$ in the latter result by using eqs.~(\ref{amu}) and (\ref{ae}), as follows,
\bea
{\rm BR}(\mu\rightarrow e \gamma)= \frac{\alpha m^3_\mu m^4_{S_1}}{ C^2_t m^2_t }\, \tau_\mu\, \bigg[\Big(\frac{a_e}{m^2_e}\Big)^2\, R^{-2} + \Big(\frac{a_\mu}{m_\mu}\Big)^2\,R^2  \bigg]
\eea
with $R=\lambda_{31}/\lambda_{32}$.  Therefore, the correlation between $(g-2)_l$'s and $\mu\to e\gamma$ is manifest.

 In Fig.~\ref{mg2}, we show the parameter space for the singlet scalar leptoquark mass, $m_{LQ}=m_{S_1}$,  and the extra leptoquark coupling $\lambda'_{32}$, where the $(g-2)_\mu$ anomaly can be explained, in green(yellow) region at $2\sigma(1\sigma)$ level. The gray region is excluded by the bound on $B(\tau\rightarrow \mu\gamma)$. We have taken $\lambda_{32}=\lambda_{33}=1(0.1)$ on left(right) plot and $\lambda'_{33}=0$. Therefore, for $m_{LQ}\lesssim 10-50\,{\rm TeV}$ under perturbativity and leptoquark couplings less than unity, the $(g-2)_\mu$ anomaly can be explained in our model, being compatible with $B(\tau\rightarrow \mu\gamma)$.

 \begin{figure}
  \begin{center}
      \includegraphics[height=0.45\textwidth]{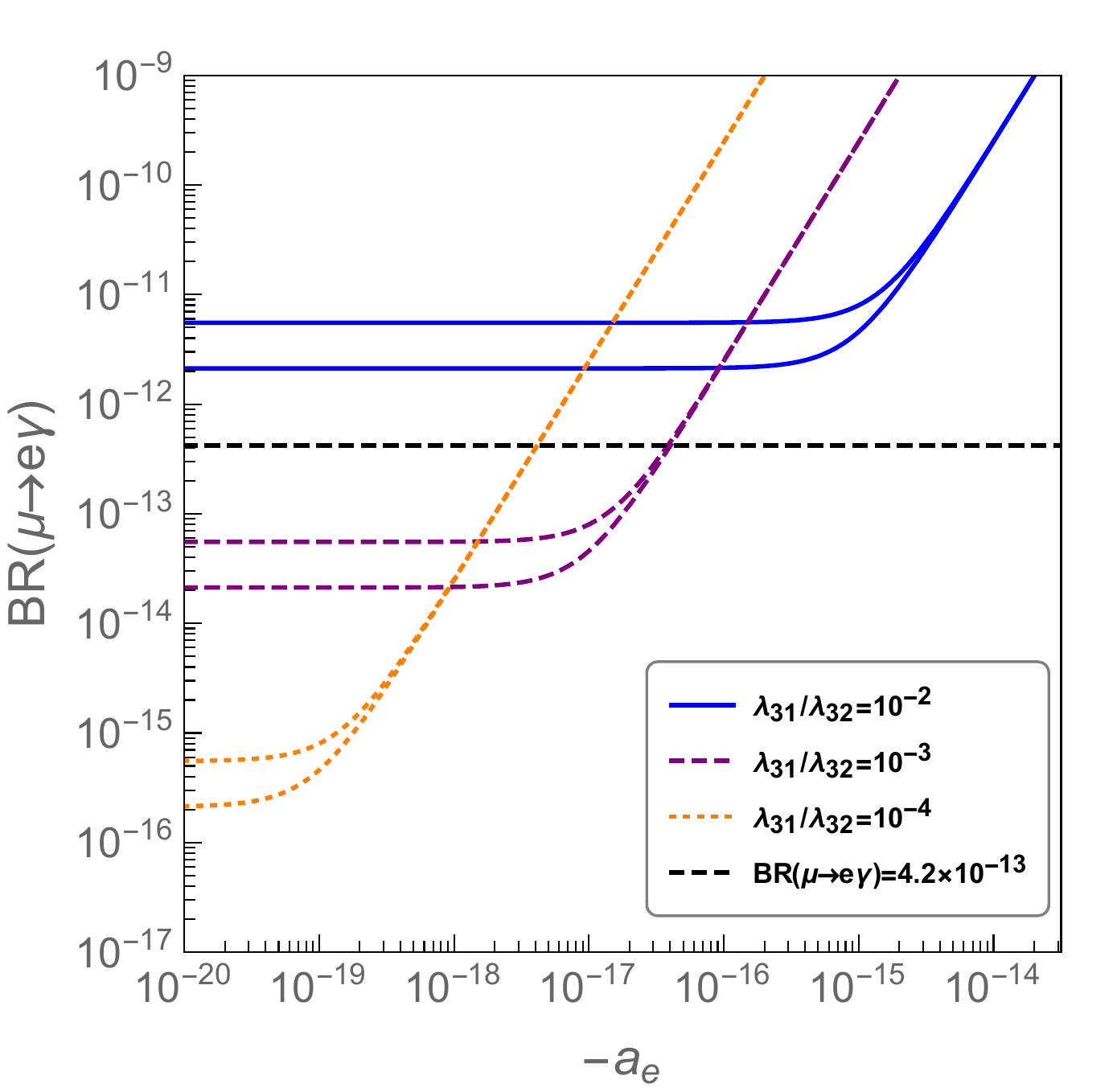}\,\,
          \includegraphics[height=0.45\textwidth]{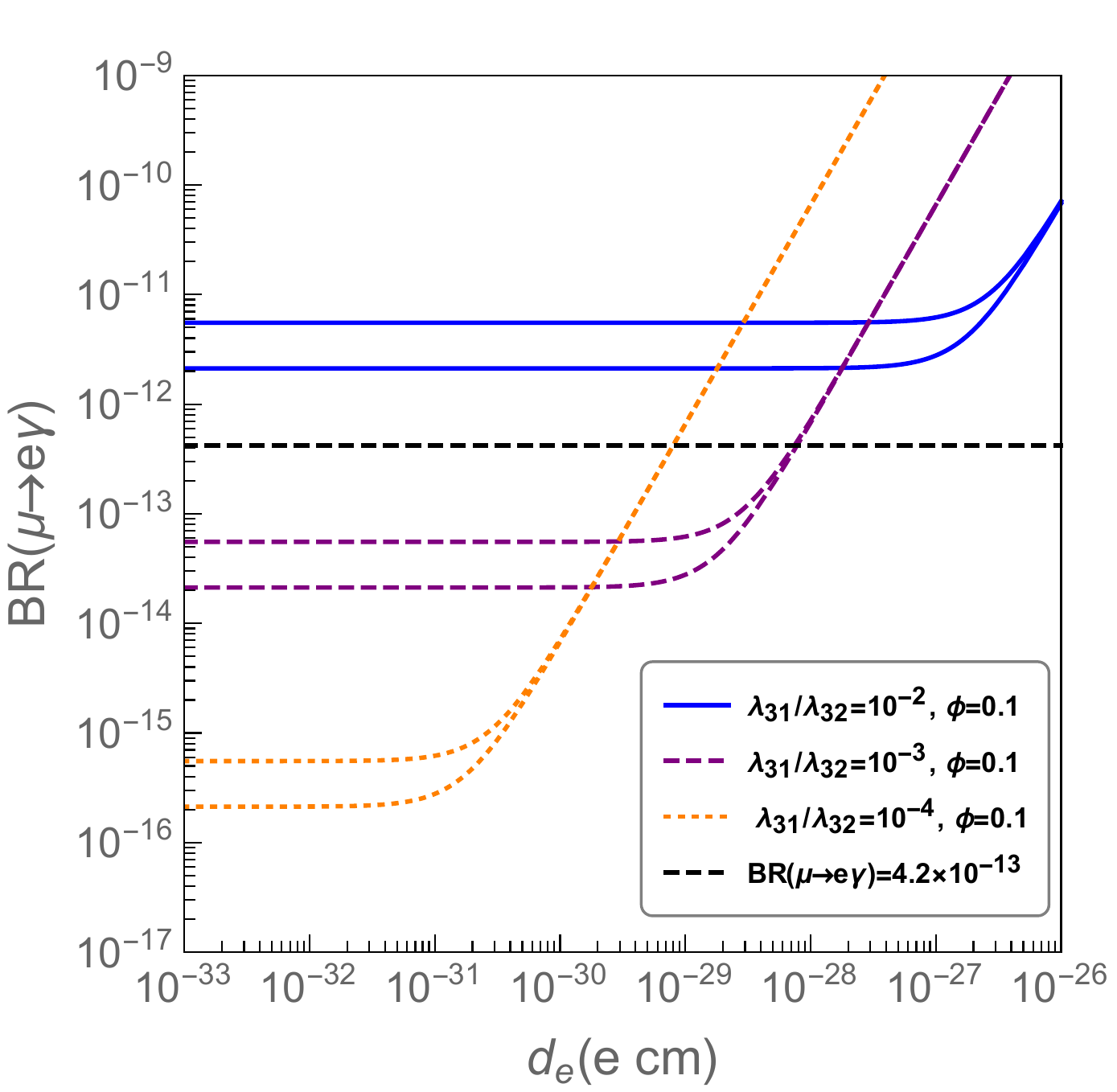}
      \end{center}
  \caption{ (Left) Correlation between ${\rm BR}(\mu\to e\gamma)$ and electron $(g-2)_e$ for the $1\sigma$ region of $(g-2)_\mu$.  (Right) Correlation between ${\rm BR}(\mu\to e\gamma)$ and electron EDM $d_e$. We took $\lambda_{31}/\lambda_{32}=10^{-4}, 10^{-3}, 10^{-2}$ from bottom to top (orange dotted, purple dashed, blue solid lines) for both plots, and the CP phase $\phi=0.1$ on right plot. The black dashed line shows the current upper limit, ${\rm BR}(\mu\to e\gamma)=4.2\times 10^{-13}$.  }
  \label{corr2}
\end{figure}

In Fig.~\ref{corr1}, we also depict the correlation between the leptonic signatures, namely,  the leptoquark contributions to the muon and electron $g-2$. We showed  the $1\sigma(2\sigma)$ bands for $(g-2)_\mu$ in green and yellow. 
We have fixed ${\rm BR}(\mu\to e\gamma)=4.2\times 10^{-13}, 10^{-15}$ on left and right plots, respectively. We varied $\lambda_{31}/\lambda_{32}=10^{-3}, 10^{-2.5}, 10^{-2}, 10^{-1.5}$ from top to bottom (black, orange, purple, blue lines) on left plot and $\lambda_{31}/\lambda_{32}=10^{-4.5}, 10^{-4}, 10^{-3.5}, 10^{-3}$ from top to bottom  (black, orange, purple, blue lines) on right plot. Then, from the left plot of Fig.~\ref{mg2}, when the bound from $B(\mu\rightarrow e\gamma)$ is saturated, the leptoquark contribution to the electron $(g-2)_e$ is smaller than about $10^{-16}$ within the $2\sigma$ region for $(g-2)_\mu$.  Requiring $B(\mu\rightarrow e \gamma)$ to be smaller than the current limit as in the right plot of  Fig.~\ref{mg2}, the electron $(g-2)_e$ becomes much smaller. 

In Fig.~\ref{corr2}, we also showed the correlation between the leptoquark contributions to ${\rm BR}(\mu\to e\gamma)$ and electron $(g-2)_e$ on left and the correlation between those to ${\rm BR}(\mu\to e\gamma)$ and electron EDM $d_e$ on right.
We took $\lambda_{31}/\lambda_{32}=10^{-4}, 10^{-3}, 10^{-2}$ from bottom to top (orange dotted, purple dashed, blue solid  lines) in both plots, and the CP phase for the electron EDM was taken to $\phi=0.1$ on right. 
Taking into account the current limit from ${\rm BR}(\mu\to e\gamma)<4.2\times 10^{-13}$, which is the region below the black dashed line in both plots, we find that the $(g-2)_e$ and the electron EDM are consequently suppressed.

\subsection{Model II: Top and charm quark loops}

We take an alternative flavor structure for leptoquark couplings as follows,
\bea
\lambda=\left(\begin{array}{ccc}  0  & 0 & 0 \\   \lambda_{21} &  0 & \lambda_{23} \\ 0 & \lambda_{32} & \lambda_{33}  \end{array} \right), \quad \kappa=\left(\begin{array}{ccc}  0  & 0 & 0 \\   \kappa_{21} &  \kappa_{22} & \kappa_{23} \\ 0 & \kappa_{32} & \kappa_{33}  \end{array} \right), \quad \lambda=\left(\begin{array}{ccc}  0  & 0 & 0 \\   \lambda'_{21} &  0 & \lambda'_{23} \\ 0 & \lambda'_{32} &  0 \end{array} \right). \label{flavor2}
\eea
The reason for the flavor structure in eq.~(\ref{flavor2}) is in order. 
First, we took $\lambda_{1i}=\lambda'_{1i}=\kappa_{1i}=0$ for satisfying the strong limits from light meson decays. As in the flavor structure in eq.~(\ref{flavor1}), we introduced the necessary leptoquark couplings, $\lambda_{33}, \lambda_{23}$ and $\kappa_{32}, \kappa_{22}$, for $B$-meson anomalies, as well as $\kappa_{23}, \kappa_{33}$ and $\lambda_{32}$ for satisfying the bound from  $B\to K\nu {\bar\nu}$, and set $\lambda'_{33}=0$ for a better fit with $C_S=C_T=0$ to $R_{D^{(*)}}$ in eq.~(\ref{RDeff}).  We need $\lambda_{32}\lambda'_{32}\neq 0$ for explaining $(g-2)_\mu$ with top-quark loops, so we took  $\lambda'_{32}=\lambda_{31}=0$ in order to avoid the strong bound from $\mu\to e\gamma$.
Similarly, we need $\lambda_{21}\lambda'_{21}\neq 0$ for explaining $(g-2)_e$ with charm-quark loops, so we took $\lambda'_{22}=\lambda_{22}=0$ for $\mu\to e\gamma$. Then, the flavor structure between the Yukawa couplings, $\lambda$ and $\lambda'$, for the singlet leptoquark, is correlated. 
Finally, for the consistency with the bounds from $B\to K\nu {\bar\nu}$, we took $\kappa_{31}=0$ for $\lambda_{31}=0$ and $\kappa_{21}\neq 0$ for $\lambda_{21}\neq 0$, subject to the condition in eq.~(\ref{lam21}).
We note that the flavor structure (\ref{flavor2}) in Model II can be realized in the basis where the mass matrices for up-quarks and charged leptons are diagonal \cite{charmquark}, as shown in eqs.~(\ref{ubasis1}) and (\ref{ubasis2}).

For the flavor structure in eq.~(\ref{flavor2}), from eq.~(\ref{g2}), the one-loop contributions with top quark and singlet leptoquark to the $(g-2)_l$ for muon and electron, are, respectively, 
\bea
a^{S_1}_\mu &=& \frac{C_t m_\mu m_t}{16\pi^2 m^2_{S_1}} \,{\rm Re}(\lambda_{32}\lambda^{\prime *}_{32}), \\
a^{S_1}_e &=& \frac{C_c m_e m_c}{16\pi^2 m^2_{S_1}} \,{\rm Re}(\lambda_{21}\lambda^{\prime *}_{21})
\eea 
with $C_c=4\log \Big(\frac{m^2_{S_1}} {m^2_{c}}\Big)-7 $. In this case, there is a top quark mass enhancement for $(g-2)_\mu$ but there is a weaker enhancement with charm quark mass for $(g-2)_e$ \cite{charmquark}. Therefore, we get the following ratio of $(g-2)_l$'s,  
\bea
\frac{a^{S_1}_e}{a^{S_1}_\mu }= \frac{C_c m_e m_c}{ C_t m_\mu m_t}\,\cdot \frac{{\rm Re}(\lambda_{21}\lambda^{\prime *}_{21}) }{{\rm Re}(\lambda_{32}\lambda^{\prime *}_{32})}.
\eea
Then, for $|a^{S_1}_e/a^{S_1}_\mu|\simeq 3\times 10^{-4} $ at the central values in the deviations in eqs.~(\ref{amu-dev}) and (\ref{ae-dev}), we would need $|{\rm Re}(\lambda_{21}\lambda^{\prime *}_{21})/{\rm Re}(\lambda_{32}\lambda^{\prime *}_{32})|\simeq 1$ for $m_{S_1}\simeq 1\,{\rm TeV}$. In this case, there is no bound from $\mu\to e\gamma$ at one-loop level.

 \begin{figure}
  \begin{center}
    \includegraphics[height=0.45\textwidth]{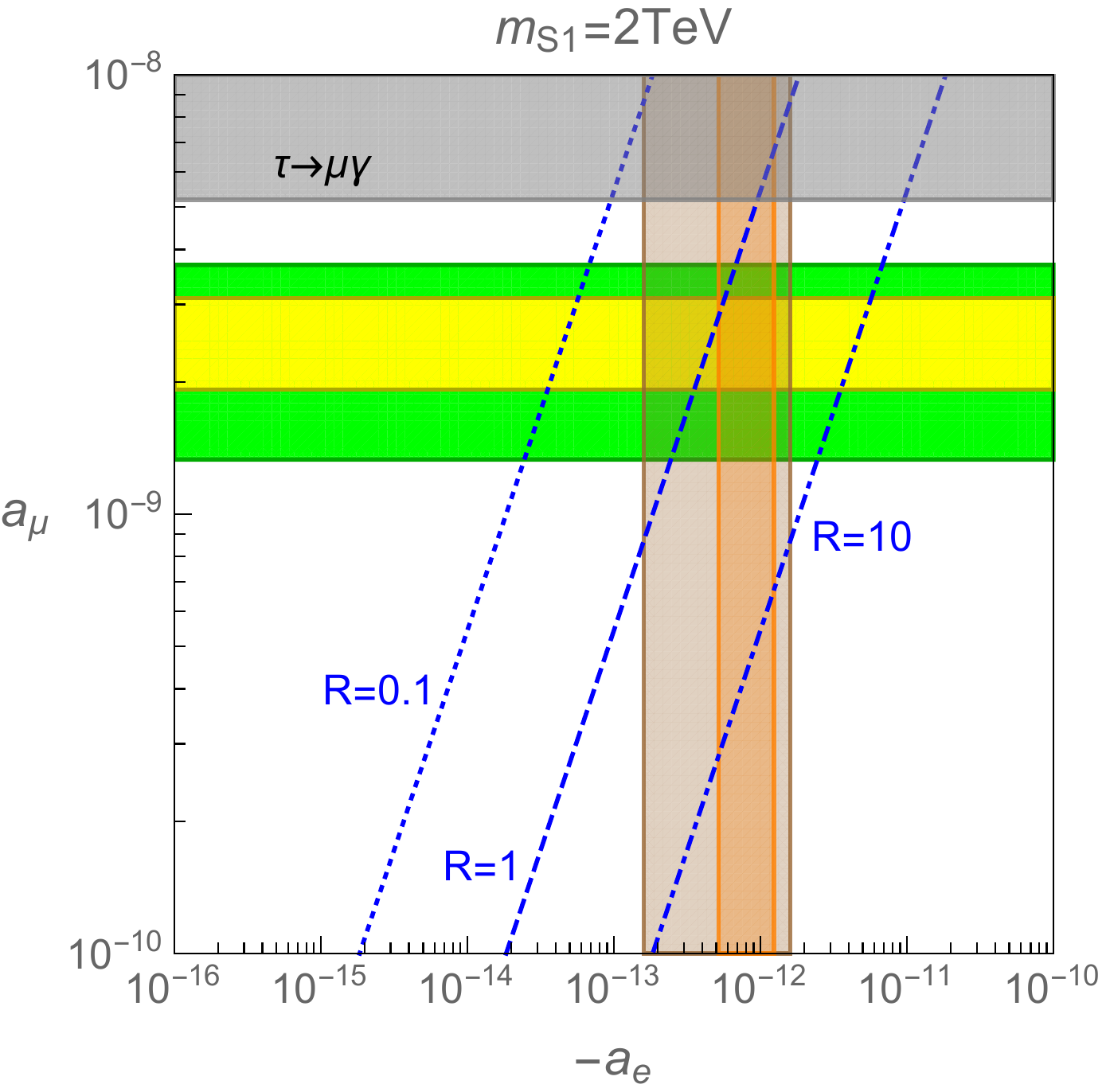} \,\,
        \includegraphics[height=0.45\textwidth]{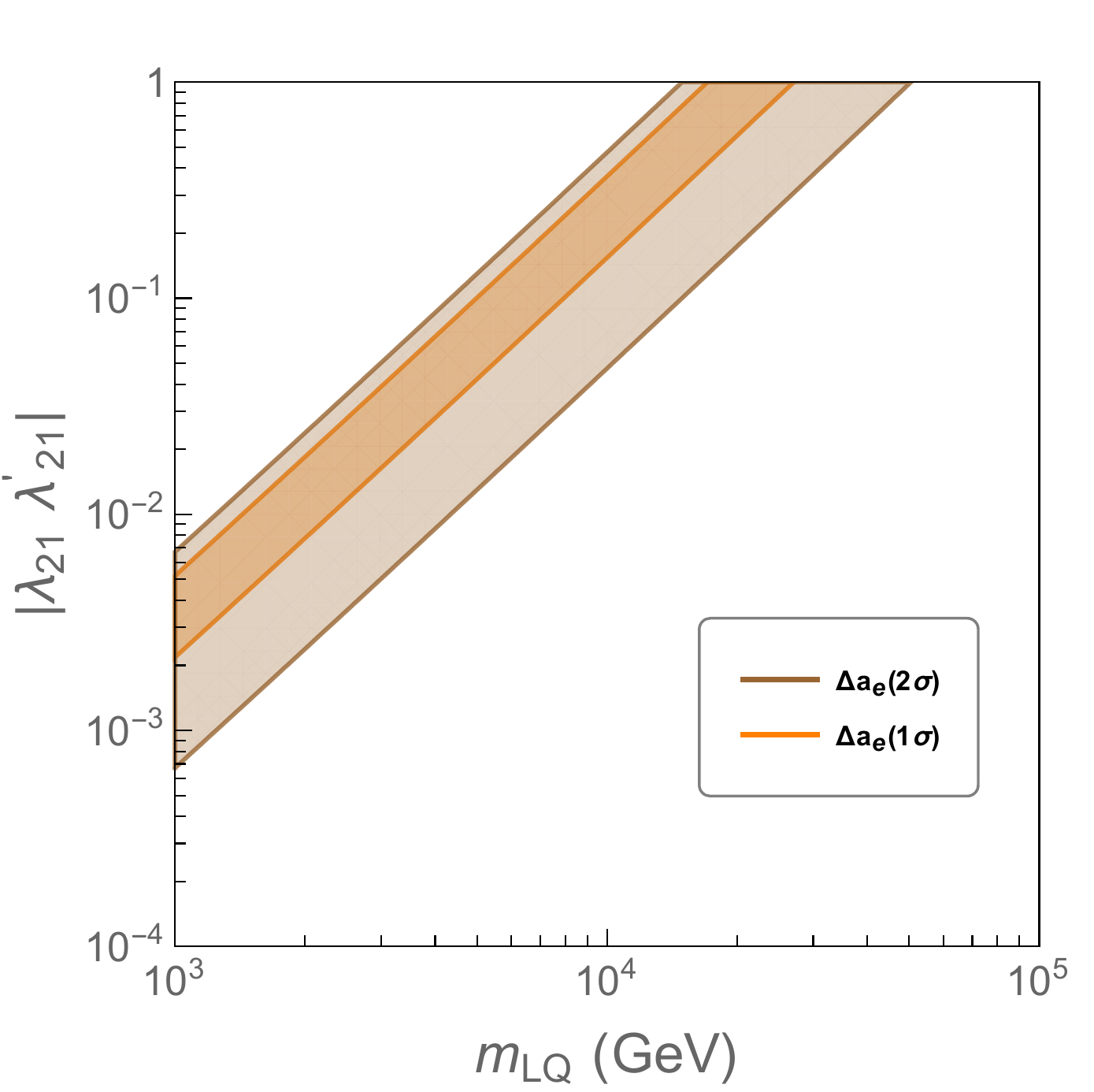}
      \end{center}
  \caption{ (Left) Correlation between muon and electron $g-2$ in Model II. We took $m_{S_1}=2\,{\rm TeV}$ and varied $R\equiv |{\rm Re}(\lambda_{21}\lambda^{\prime *}_{21})/{\rm Re}(\lambda_{32}\lambda^{\prime *}_{32})|=0.1,1, 10$ in dotted, dashed and dotdashed lines, respectively. We also indicated the bound from ${\rm BR}(\tau\to \mu\gamma)$ in gray for $\lambda_{33}/\lambda_{32}=25$. (Right) Parameter space for $m_{LQ}=m_{S_1}$ and $|\lambda_{21}\lambda^{\prime }_{21}|$ allowed by $(g-2)_e$ in Model II. Green(yellow) regions show the $2\sigma(1\sigma)$ bands for  $(g-2)_\mu$ and brown(orange) regions show the $2\sigma(1\sigma)$ bands for  $(g-2)_e$.  }
  \label{model2}
\end{figure}

In Fig.~(\ref{model2}), we draw the correlation between  muon and electron $g-2$ in Model II on left and the parameter space for $m_{LQ}=m_{S_1}$ and $(|\lambda_{21}\lambda^{\prime }_{21}|)^{1/2}$ explaining the $(g-2)_e$ anomaly in Model II on right. In the left plot of Fig.~(\ref{model2}), we took $m_{S_1}=2\,{\rm TeV}$ and several values of $R\equiv |{\rm Re}(\lambda_{21}\lambda^{\prime *}_{21})/{\rm Re}(\lambda_{32}\lambda^{\prime *}_{32})|=0.1,1, 10$ in dotted, dashed and dotdashed lines, respectively. The simultaneous explanation of $(g-2)_\mu$ and $(g-2)_e$ anomalies is possible for $R\simeq 1$, which means that $|\lambda_{21}\lambda^{\prime }_{21}|\sim|\lambda_{32}\lambda^{\prime }_{32}|\gtrsim 10^{-3}$ for $m_{LQ}\gtrsim 1\,{\rm TeV}$.

Moreover, from eq.~(\ref{g2}), the leptoquark contributions to the lepton EDMs are given by
\bea
d_\mu &=&  \frac{C_t m_t}{32\pi^2 m^2_{S_1}} \,{\rm Im}(\lambda_{32}\lambda^{\prime *}_{32}), \\
d_e &=& \frac{C_c  m_c}{32\pi^2 m^2_{S_1}} \,{\rm Im}(\lambda_{21}\lambda^{\prime *}_{21}).
\eea
Then, we find that the electron EDM is related to the electron $(g-2)_e$ and the CP phase by
\bea
d_e\simeq  (10^{-29}\,{\rm e\,cm}) \bigg(\frac{\phi}{6\times 10^{-6}} \bigg) \bigg(\frac{a_e}{9\times 10^{-13}} \bigg)
\eea
with $\phi={\rm Im}(\lambda_{21}\lambda^{\prime *}_{21})/{\rm Re}(\lambda_{21}\lambda^{\prime *}_{21})$. Therefore, the CP phase in the leptoquark couplings are severely constrained to be less than $6\times 10^{-6}$ in the region where the electron $(g-2)_e$ anomaly is explained. 
On the other hand, from eqs.~(\ref{BRtaumu}) and (\ref{BRtaue}), we obtain the same branching ratios for $\tau\to \mu\gamma$ and $\tau\to e\gamma$ as in Model I, so it constrains the model equally as in Model I. However, from eq.~(\ref{mue}), the new contribution to the branching ratio for $\mu\to e\gamma$ vanishes at one-loop, so it is possible to reconcile the strong tension between the limit from $\mu\to e\gamma$ and  the simultaneous explanation of $(g-2)_\mu$ and $(g-2)_e$ anomalies.

\subsection{Neutrino masses from leptoquarks}

We introduce a doublet leptoquark $S_2$ with hypercharge $Y=+\frac{1}{6}$, with the following interactions to the SM fermions and the singlet leptoquark,
\bea
{\cal L}_{S_2} = -{\tilde\lambda}_{ij} {\bar d}_{Ri} l_{Lj} S_2 - \mu\, {\tilde H} S_2 S_1 +{\rm h.c.}
\eea
where $\mu$ is the dimensionful parameter and ${\tilde H}=i\sigma^2 H^*$ is the complex conjugate of the SM Higgs doublet.  Then,  the lepton number is violated due to the leptoquark couplings and there appears a mixing between the singlet and doublet leptoquarks after electroweak symmetry breaking \footnote{We note that there can be a nonzero mixing between the singlet and triplet leptoquarks in our model by ${\cal L}_{\rm mix}=-\lambda_m H^\dagger \Phi H S^*_1+{\rm h.c.}$ In this case, however, there is no lepton number violation at the renormalizable level, so neutrino masses are absent. }. Consequently, neutrino masses can be generated due to loop corrections with leptoquarks, similarly as in R-parity violating supersymmetry \cite{neutrinomass}, and the dominant contributions to the neutrino mass matrix stem from the top quark at one-loop level, given for $m_{S_2}\gg m_{S_1}$ by
\bea
(m_\nu)_{ij} =A\,(\lambda_{31},\lambda_{32},\lambda_{33}) \left(\begin{array}{c} {\tilde\lambda}^*_{31} \\ {\tilde\lambda}^*_{32} \\ {\tilde\lambda}^*_{33} \end{array} \right)
\eea
where
\bea
A\simeq \frac{v \, m_t \mu}{8\sqrt{2} \pi^2 m^2_{S_1}}
\eea
and $v=246\,{\rm GeV}$. 
The above mass matrix has unit rank, so there is only one nonzero mass eigenvalue, $m_{\nu,3}=A(\lambda_{31} {\tilde\lambda}^*_{31}+\lambda_{32} {\tilde\lambda}^*_{32}+\lambda_{33} {\tilde\lambda}^*_{33})$.
Therefore, we need to introduce at least one more doublet leptoquark to realize the current neutrino oscillation data.
Here, we can set an upper bound on the leptoquark couplings from $m_\nu<0.1\,{\rm eV}$, as follows,
\bea
{\rm max}\Big(|\lambda_{3i}{\tilde\lambda}^*_{3i}|\Big)<10^{-9} \bigg(\frac{v}{\mu} \bigg) \bigg(\frac{m_{S_1}}{1\,{\rm TeV}} \bigg)^2.
\eea
As a result, it is enough to introduce tiny extra couplings for the doublet leptoquark to get neutrino masses, so the additional contributions from the doublet leptoquark to $b\rightarrow s\mu^+ \mu^-$ (namely, with the Wilson coefficients, $C^{\prime \mu}_9=-C^{\prime \mu}_{10}$, in eq.~(\ref{RKeff})) \cite{LQ-hmlee} are suppressed.

\section{Conclusions}

We revisited the models of scalar leptoquarks in light of the $B$-meson anomalies in the quark sector and the $(g-2)_l$ anomalies in the lepton sector. Then, we determined the flavor structure for the leptoquark couplings phenomenologically, based on the bounds from other $B$-meson decays and flavor-violating decays of leptons. 

First, we showed that there is a strong correlation between the $\mu\to e\gamma$ and the magnetic or electric dipole moments of electron in the minimal scenario for the $B$-meson anomalies and the $(g-2)_\mu$ anomaly.  In this case, there are  top-quark mass enhancements in the magnetic dipole moments of muon and electron, so there is a strong bound from the flavor violation in $\mu\to e\gamma$. 
On the other hand, in the benchmark model oriented to accommodate  both $(g-2)_l$ anomalies,  the magnetic dipole moment of muon still has the top-quark mass enhancement, but the magnetic dipole moment of electron is relatively suppressed by the charm-quark mass. Therefore, we can evade the bounds from  $\mu\to e\gamma$ and other flavor violating decays of leptons and there is an interesting relation between the leptoquark contributions to the $(g-2)_\mu$ and $(g-2)_e$ values. We also found that neutrino masses can be generated from the top-quark loops in the model augmented with several doublet leptoquarks, being consistent with the $B$-meson decays.

\acknowledgments

The work is supported in part by Basic Science Research Program through the National Research Foundation of Korea (NRF) funded by the Ministry of Education, Science and Technology (NRF-2019R1A2C2003738 and NRF-2021R1A4A2001897).


\end{document}